\documentclass[useAMS,usenatbib]{mn2e}
\newcommand{\bib}{\bibitem[\protect\citeauthoryear}
\usepackage[dvips]{graphicx}
\usepackage[T1]{fontenc}
\usepackage{ae,aecompl}

\title[NGC\,315]{Multifrequency observations of the jets in the radio galaxy
  NGC\,315} \author[R.A. Laing et al.]{R.A. Laing\thanks{E-mail:
  rlaing@eso.org}$^{1}$, J.R. Canvin$^{2,3}$, W.D. Cotton$^{4}$,
  A.H. Bridle$^4$\\$^1$ European Southern Observatory,
  Karl-Schwarzschild-Stra\ss e 2, D-85748 Garching-bei-M\"{u}nchen, 
Germany\\$^2$ School of Physics, University of Sydney, A28, Sydney,
  NSW 2006, Australia\\$^3$ University of Oxford, Department of Astrophysics, 
  Denys Wilkinson
  Building, Keble Road, Oxford OX1 3RH \\$^4$ National Radio Astronomy
  Observatory, 520 Edgemont Road, Charlottesville, VA 22903-2475, U.S.A.}

\date{Received}

\pagerange{\pageref{firstpage}--\pageref{lastpage}} \pubyear{2005}

\begin{document}
\label{firstpage}

\maketitle

\begin{abstract}
We present images of the jets in the nearby radio galaxy NGC\,315 made with the
VLA at five frequencies between 1.365 and 5\,GHz with resolutions between 1.5 and
45\,arcsec FWHM. Within 15\,arcsec of the nucleus, the spectral index of the
jets is $\alpha = 0.61$. Further from the nucleus, the spectrum is flatter, with
significant transverse structure. Between 15 and 70\,arcsec from the nucleus,
the spectral index varies from $\approx$0.55 on-axis to $\approx$0.44 at the
edge. This spectral structure suggests a change of dominant particle
acceleration mechanism with distance from the nucleus and the transverse
gradient may be associated with shear in the jet velocity field. Further from
the nucleus, the spectral index has a constant value of 0.47. We derive the
distribution of Faraday rotation over the inner $\pm$400\,arcsec of the radio source
and show that it has three components: a constant term, a linear gradient (both
probably due to our Galaxy) and residual fluctuations at the level of 1 --
2~rad\,m$^{-2}$. These residual fluctuations are smaller in the brighter
(approaching) jet, consistent with the idea that they are produced by magnetic
fields in a halo of hot plasma that surrounds the radio source. We model this
halo, deriving a core radius of $\approx$225\,arcsec and constraining its
central density and magnetic-field strength. We also image the apparent
magnetic-field structure over the first $\pm$200\,arcsec from the nucleus.
\end{abstract}

\begin{keywords}
galaxies: jets -- radio continuum:galaxies -- galaxies: individual: NGC\,315 --
magnetic fields -- polarization -- MHD
\end{keywords}

\section{Introduction}
\label{intro}

The giant FR\,I \citep{FR74} radio source NGC\,315 was first imaged by
\citet{Brid76}, who showed that it has an angular size of nearly 1$^\circ$.  The
extended radio structure is described in more detail by \citet{Brid79},
\citet{Fom80}, \citet{Willis81}, \citet{Jaegers}, \citet{Venturi93},
\citet{Mack97} and \citet{Mack98}. The main jet has also been imaged extensively
on parsec scales \citep{Linfield81,Venturi93,Cotton99,Xu00}. X-ray emission from
the first 10\,arcsec of the main jet was detected by \citet{WBH}, but no optical
emission from this region has yet been reported.

The source is associated with a giant elliptical galaxy at a redshift of 0.01648
\citep{Trager}, giving a scale of 0.335\,kpc/arcsec for our adopted cosmology
(Hubble constant $H_0$ = 70\,$\rm{km\,s^{-1}\,Mpc^{-1}}$, $\Omega_\Lambda =
0.7$ and $\Omega_M = 0.3$).  NGC\,315 is a member of a group or poor cluster of
galaxies \citep{Nolthenius,Miller02} located in one of the filaments of the
Pisces-Perseus supercluster \citep{Ensslin01,Huchra}. HST images
\citep{Verdoes99} show a 2.5-arcsec diameter dust lane and a nuclear
point source. The dust lane is associated with a disk of ionized gas which is
probably in ordered rotation \citep{Noel-Storr}. CO emission, also with a line
profile indicating rotation, was detected by \citet{Leon}. The inferred mass of
molecular hydrogen is $(3.0 \pm 0.3) \times 10^8$\,M$_\odot$ and the cold gas is
likely to be cospatial with the dust. HI absorption against the
nucleus was detected by \citet{vanG89}. There is evidence for a weak, polarized
broad H$\alpha$ line in the nuclear spectrum \citep{Ho97,Barth99,Noel-Storr}.
Hot gas associated with the galaxy has been imaged using {\sl ROSAT} and {\sl
Chandra} \citep{WB,WBH}.

Within $\approx$\,90 arcsec of the nucleus, the jets in NGC\,315 are initially
narrow, then expand rapidly (``flare'') and re-collimate \citep{Brid82,CLBC}. We
have modelled the inner $\pm$70\,arcsec of this {\em flaring region} as a
two-sided, symmetrical, relativistic flow, fitting to deep, high-resolution VLA
observations at 5\,GHz in order to derive the three-dimensional distributions of
velocity, proper emissivity and magnetic-field structure \citep{CLBC}. Our main
conclusions are as follows.
\begin{enumerate}
\item The jets are inclined by 
$38^\circ \pm 2^\circ$ to the line of sight. 
\item Where they first brighten, their on-axis velocity is $\beta = v/c \approx
0.9$. They decelerate to $\beta \approx 0.4$ between 8 and 18\,kpc from the
nucleus (15 -- 33\,arcsec in projection) and the velocity thereafter remains
constant.
\item The ratio of the speed at the edge of the jet to its value on-axis ranges
from $\approx$0.8 close to the nucleus to $\approx$0.6 further out.
\item The longitudinal profile of proper emissivity 
is split into three power-law regions separated by shorter transition zones and
the emission is intrinsically centre-brightened.
\item  To a first approximation, the magnetic field evolves from a mixture of
longitudinal and toroidal components to predominantly toroidal by 26\,kpc
(48\,arcsec in projection).
\item Simple adiabatic models fail to fit the emissivity
variations.  
\end{enumerate}

In the present paper, we investigate the energy spectrum of the relativistic
particles in the jets of NGC\,315 in the context of the models developed by
\citet{CLBC}. We use VLA observations at frequencies between 1.365 and
5\,GHz\footnote{The 5-GHz observations are those discussed by \citet{CLBC}} to
derive the spectrum of the jets at resolutions of 5.5 and 1.5\,arcsec and relate
the observed spectral gradients to velocity, emissivity and field structure. A
separate paper (Worrall et al., in preparation) will describe the radio structure of
the main jet at high resolution and its relation to new {\sl Chandra} images.

We also determine the variations of Faraday rotation over the jets and test the
hypothesis that these result from magnetic-field irregularities in hot, X-ray
emitting plasma associated with the surrounding group of galaxies. Finally, we
determine the apparent magnetic-field structure of the jets on scales larger
than those covered by \citet{CLBC}.

In Section~\ref{obs}, we describe the observations and their reduction. The
total-intensity images are presented in Section~\ref{Images} and we use them to
derive distributions of spectral index in Section~\ref{Spectra}. We then discuss
the distributions of Faraday rotation (Section~\ref{Faraday}) and apparent
magnetic-field structure (Section~\ref{Field}) derived from observations of
linear polarization.  Section~\ref{Summary} summarizes our main results.

\section{Observations and images}
\label{obs}

\subsection{Observations}

VLA data were obtained at 4.985\,GHz in the B, C/D and A/D configurations as
described by \citet{Venturi93} and \citet{Cotton99}. These were supplemented by
additional observations in the A and C configurations with a centre frequency of
4.860\,GHz to give complete coverage of the spatial scales accessible to the VLA
in a single pointing.  In order to map Faraday rotation, we observed at 1.365,
1.413, 1.485 and 1.665\,GHz in the B and C configurations of the VLA using a
lower bandwidth. We also extracted observations in A configuration at 1.413\,GHz
from the VLA archive.\footnote{In addition, we re-analysed the C-configuration
dataset at 8.4\,GHz from \citet{Venturi93}, but poor weather during the
observations precluded accurate absolute flux and polarization calibration, so
we do not discuss them here.} A journal of observations is given in
Table~\ref{record}.

\begin{center}
\begin{table}
\caption{Record of VLA observations. $\nu$ and $\Delta\nu$ are the centre
  frequency and bandwidth, respectively, and t is the on-source integration
  time.}
\begin{center}
\begin{tabular}{clllc}
\hline
 Config- &    Date     & $\nu$ & $\Delta\nu$  & t \\
 uration &             & (MHz)     & (MHz)      & (min)  \\
\hline
          B   & 1989 Apr 13 & 4985.1  & 50  & 279  \\
          B   & 1995 Oct 25 & 4985.1  & 50  & 396   \\
          C/D & 1996 May 10 & 4985.1  & 50  & 417    \\
          A/D & 1996 Oct 07 & 4985.1  & 50  & 428  \\
          A   & 1996 Nov 2  & 4860.1  & 100 & 586 \\
          C   & 1997 Jul 12 & 4860.1  & 100 & 283 \\
          A   & 1980 Dec 21 & 1413.0  & 25  & 473   \\
          B   & 2001 Mar 19 & 1365.0  & 12.5 & 108 \\
          B   & 2001 Mar 19 & 1413.0  & 12.5 & 108 \\
          B   & 2001 Mar 19 & 1485.0  & 12.5 & 109 \\
          B   & 2001 Mar 19 & 1665.0  & 12.5 & 107 \\
          C   & 2001 Jul 17 & 1365.0  & 12.5 &  61 \\
          C   & 2001 Jul 17 & 1413.0  & 12.5 &  61 \\
          C   & 2001 Jul 17 & 1485.0  & 12.5 &  65 \\
          C   & 2001 Jul 17 & 1665.0  & 12.5 &  57 \\
\hline 
\end{tabular}
\end{center}
\label{record}
\end{table}
\end{center}

\begin{figure}
\includegraphics[width=8.5cm]{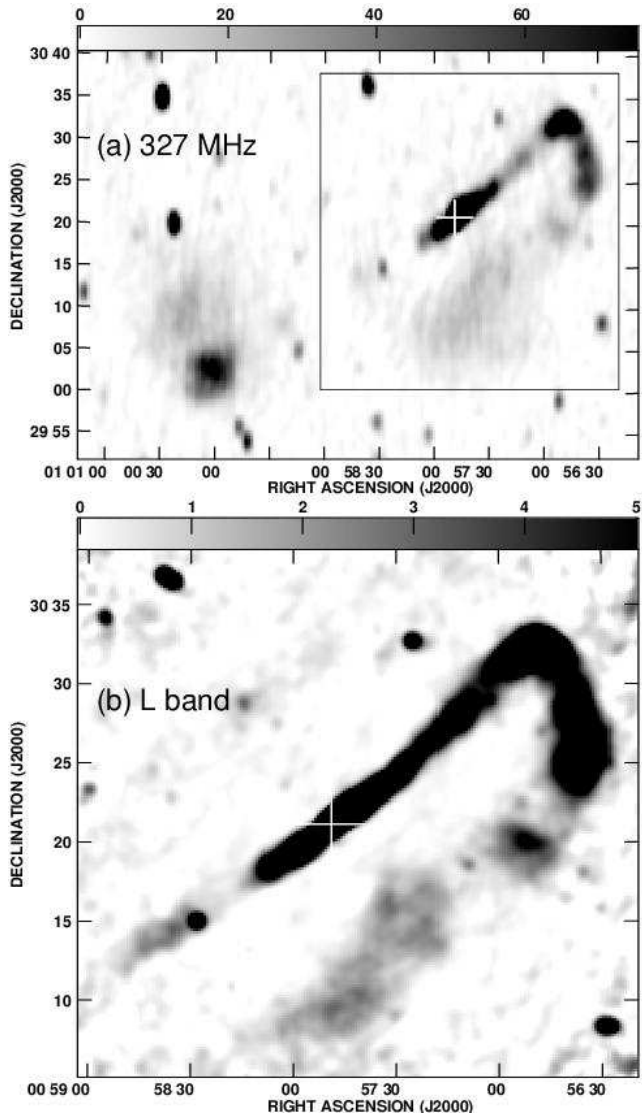}
\caption{Grey-scales of total intensity at low resolution. (a) WSRT image at
  327\,MHz, showing the full extent of the radio emission from NGC\,315
  \citep{Mack97}. The restoring beam is 110 $\times$ 55\,arcsec$^2$ FWHM in
  position angle 0 and the grey-scale range is 0 -- 75\,mJy (beam area)$^{-1}$.
  The box marks the area covered by panel (b) and Fig.~\ref{fig:i5.5full}. (b)
  The mean of all four VLA L-band images, tapered to a resolution of 45\,arcsec
  FWHM and corrected for attenuation by the primary beam before averaging.  The
  grey-scale range, 0 -- 5\,mJy (beam area)$^{-1}$, has been chosen to emphasise
  the diffuse emission. In both panels, the cross marks the position of the
  core.
\label{fig:ilow}}
\end{figure}

\subsection{Data reduction}
\label{reduction}

All of the data reduction was done in the {\sc aips} package.  Initial amplitude
and phase calibration were applied using standard methods and the flux-density
scales were set using observations of 3C\,48 or 3C\,286.  Standard instrumental
polarization calibration was applied and the zero-points of ${\bf E}$-vector
position angle were determined using observations of 3C\,138 or 3C\,286. The
data for each configuration were first adjusted to a common phase centre in
J2000 coordinates, imaged and self-calibrated separately.  They were then
concatenated in turn, starting with the widest configuration.  The slight
difference in centre frequency between the datasets at 4.86 and 4.985\,GHz was
ignored (we show in Section~\ref{RM} that this has a negligible effect on the
analysis of polarization) and we will refer to the combination as the ``5\,GHz
dataset''. At this frequency, the core showed significant variability between
observations (cf.\ \citealt{Lazio01}) and the flux density of the unresolved
component in the larger-configuration dataset was adjusted to match that
observed with the smaller configuration when both were imaged at matched
resolution. No core variability was detected at lower frequencies. A further
iteration of phase self-calibration was done after each combination.  Our final
datasets are listed in Table~\ref{Datasets}, together with the minimum and
maximum spatial scales they sample.

\begin{center}
\begin{table}
\caption{Final uv datasets. The columns are: (1) centre frequency, (2) array
  configurations used, (3) minimum and (4) maximum spatial scales.
\label{Datasets}}
\begin{tabular}{llll}
\hline
$\nu$ & Configurations &\multicolumn{2}{c}{Scales (arcsec)}\\
(MHz) &    & Min  &  Max \\
1365 & BC        & 4   & 900\\
1413 & ABC       & 1.2 & 900\\
1485 & BC        & 4   & 900\\
1665 & BC        & 4   & 900\\
4985/4860 & ABCD & 0.4 & 300\\
\hline
\end{tabular}
\end{table}
\end{center}

Our observations were designed to give the maximum sensitivity for the inner
jets of NGC\,315 and were therefore taken with the pointing centre on or near
the nucleus. As can be seen from Table~\ref{Datasets}, the maximum spatial scale
sampled adequately at 5\,GHz is $\approx$300\,arcsec. Even at L-band, the lobe
associated with the counter-jet (\citealt{Mack97} and Fig.~\ref{fig:ilow}a) is
severely attenuated by the primary-beam response of the VLA and is not visible
on our images. We did not recover the total flux density of the source at any
frequency, so we estimated appropriate zero-spacing flux densities from
the shortest-spacing visibility amplitudes. We made images at five resolutions:
45, 5.5, 2.35, 1.5 and 0.4 arcsec FWHM, using similar baseline ranges at all
frequencies and weighting the data in the uv plane as required.  After imaging,
we made both {\sc clean} and maximum-entropy deconvolutions. Although the latter
algorithm gave slightly smoother images, it introduced a significant
large-scale ripple parallel to the jet axis, whereas {\sc clean} gave a flat
background. We therefore show the {\sc clean} images, although we quote
quantitative results only where the two deconvolution methods agree. We also
compared the $I$ images made with and without zero-spacing flux densities and
before and after subtraction of a local zero-level. None of these differences
led to significant changes in spectral index or degree of polarization compared
with the errors quoted below.  After deconvolution, all of the images were
corrected for primary beam attenuation. We then took averages of the $I$ images
at 1.365 -- 1.665\,MHz (``mean L-band images'').

Data in Stokes $Q$ and $U$ were imaged without zero-spacing flux densities and
{\sc clean}ed.  A first-order correction for Ricean bias \citep{WK} was applied
to the images of polarized intensity $P = (Q^2+U^2)^{1/2}$ used to derive the
degree of polarization $p = P/I$. 

The off-source noise levels at the centre of the field for the final images are
given in Table~\ref{noise} (the 0.4-arcsec image at 5\,GHz is discussed in
detail by Worrall et al., in preparation, and is therefore not considered
further here). Note that the wide-field L-band images at a resolution of
5.5\,arcsec are significantly affected by bandwidth smearing in their outer
regions, images of point sources being broadened by a factor of 2 in the radial
direction at a distance of 22\,arcmin from the phase centre \citep{ObsSS}.  This
limitation needs to be taken into account only in the discussion of the source
morphology on large scales (Section~\ref{Images}). Measurements of spectra are
restricted to the inner 200\,arcsec of the field, where the effects of bandwidth
smearing are $<$3\% in peak intensity or image size for any of our
frequency/resolution combinations. Our estimates of Faraday rotation, which
extend to larger scales, should not be affected systematically by bandwidth
smearing. 

\begin{center}
\begin{table}
\caption{Image resolutions and noise levels. $\sigma_I$ is the
  off-source noise level on the $I$ image; $\sigma_P$ the average of
  the noise levels for $Q$ and $U$. The noise levels were evaluated before
  correction for the primary beam response and apply only at the centre of the
  field for corrected images. Asterisks denote the images used by \citet{CLBC}
\label{noise}}
\begin{tabular}{llcc}
\hline
$\nu$ & FWHM &\multicolumn{2}{c|}{rms noise level} \\
(GHz)     &  (arcsec)  &\multicolumn{2}{c|}{($\mu$Jy / beam area)} \\
          &  &$\sigma_I$&$\sigma_P$ \\
Mean L & 45 & 200  &$-$    \\
1.365 & 5.5 &41 & 35  \\
1.413 & 5.5 &38 & 34  \\
1.485 & 5.5 &37 & 33  \\
1.665 & 5.5 &37 & 30  \\
Mean L & 5.5 & 17 &$-$  \\
5 & 5.5 &15 & 12  \\
5*& 2.35 & 10 & 8 \\
1.413 & 1.5 & 36 & 36 \\
5 & 1.5 & 10 & 10 \\
5*& 0.40 & 13 & 7 \\
\hline
\end{tabular}
\end{table}
\end{center}


\section{Total intensity}
\label{Images}

\begin{figure*}
\includegraphics[width=17cm]{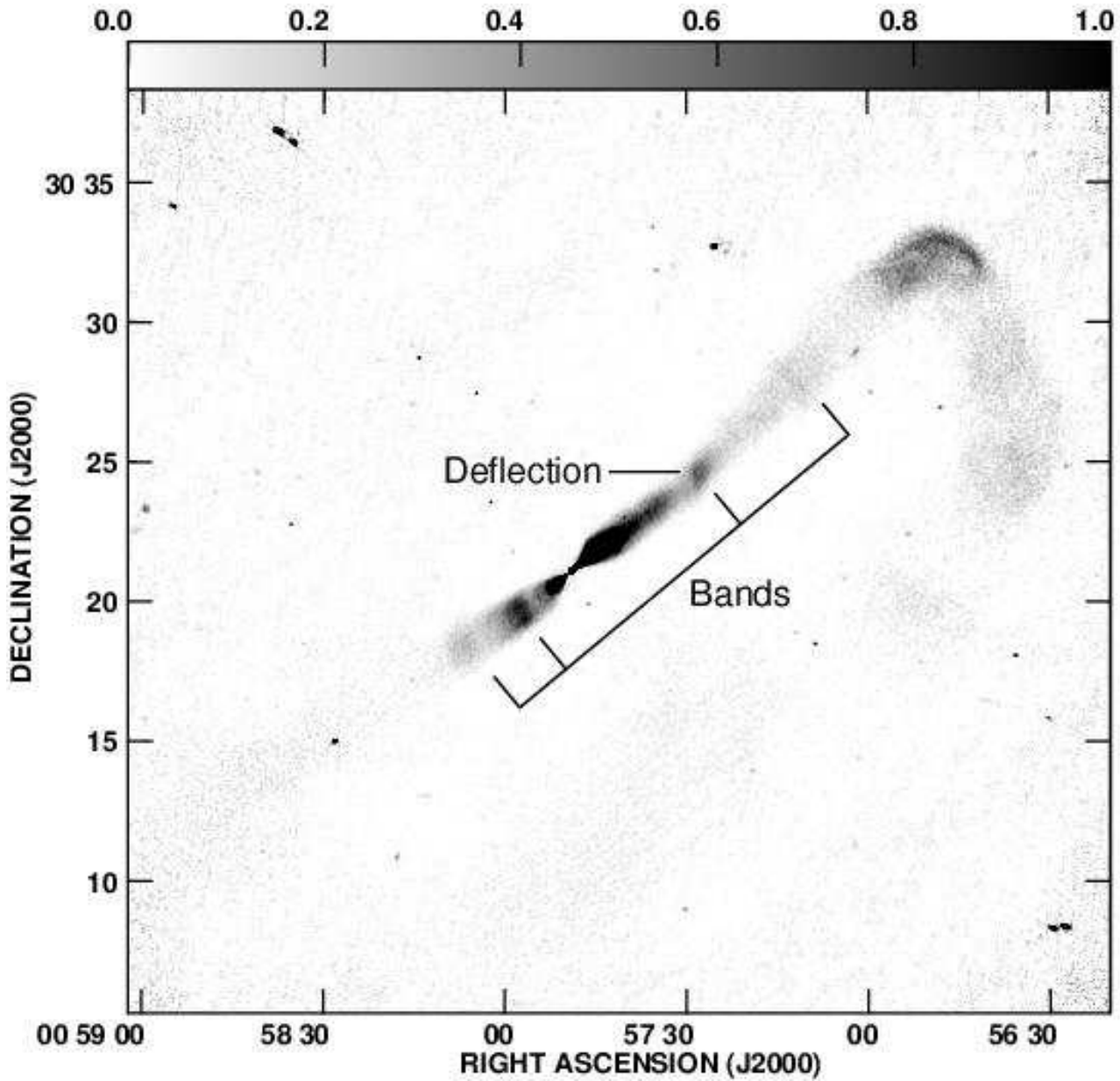}
\caption{Grey-scale of total intensity from the mean of all four L-band images,
   corrected for attenuation by the primary beam before averaging. The
   resolution is 5.5\,arcsec FWHM and the grey-scale range, 0 -- 1\,mJy
   (beam area)$^{-1}$, is marked by the labelled wedge. The area is the same as
   in Fig.~\ref{fig:ilow}(b). Note that the effects of bandwidth smearing are
   significant in this image (see Section~\ref{reduction}).
\label{fig:i5.5full}}
\end{figure*}

\begin{figure}
\includegraphics[width=8.5cm]{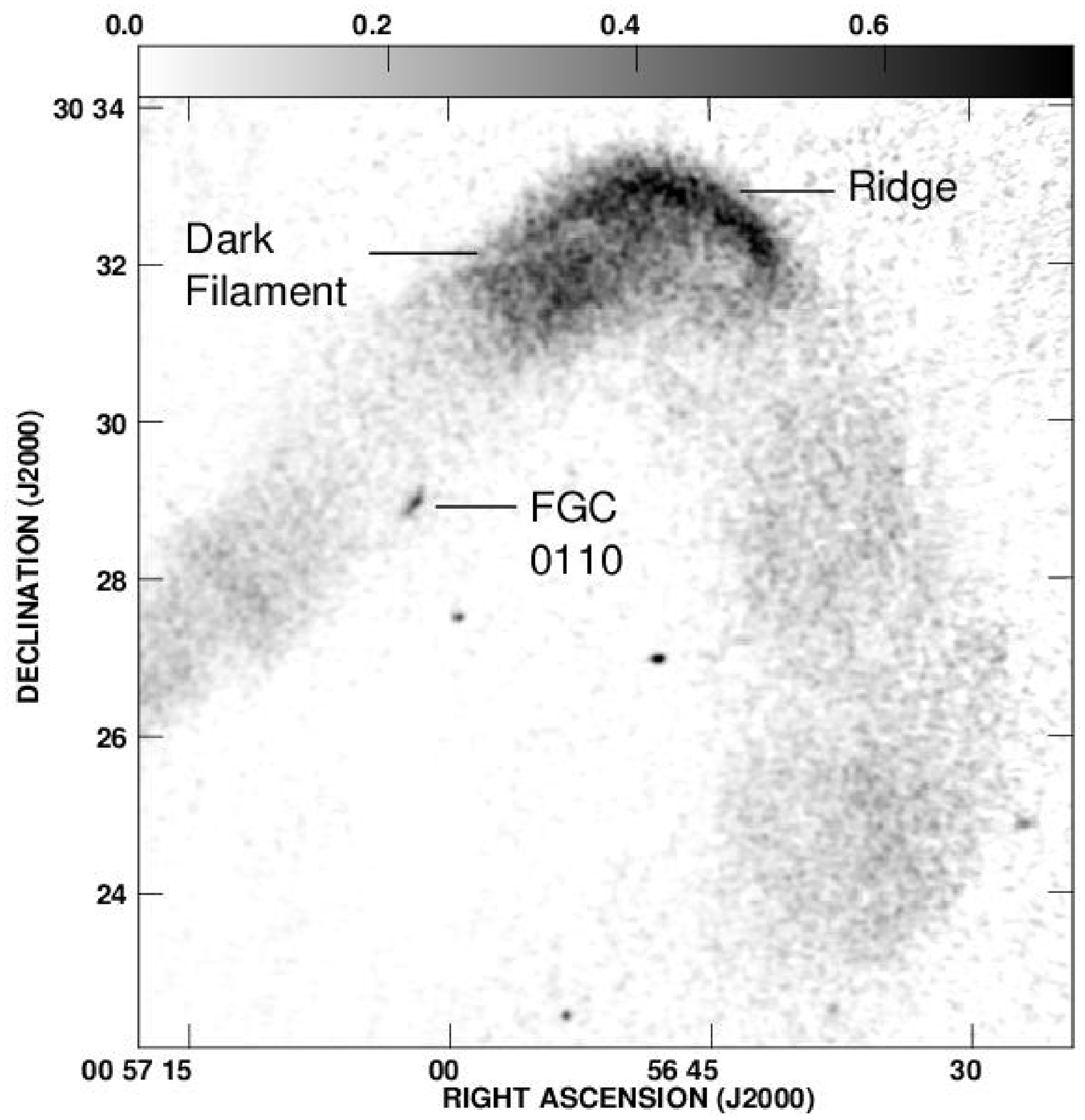}
\caption{Grey-scale of total intensity from the mean of all four L-band images,
   corrected for attenuation by the primary beam before averaging and covering
   the area around the sharp bend in the main jet. The resolution is 5.5\,arcsec
   FWHM and the  grey-scale range is  0 -- 0.75\,mJy (beam area)$^{-1}$.
\label{fig:ibend}}
\end{figure}

\begin{figure*}
\includegraphics[width=14cm]{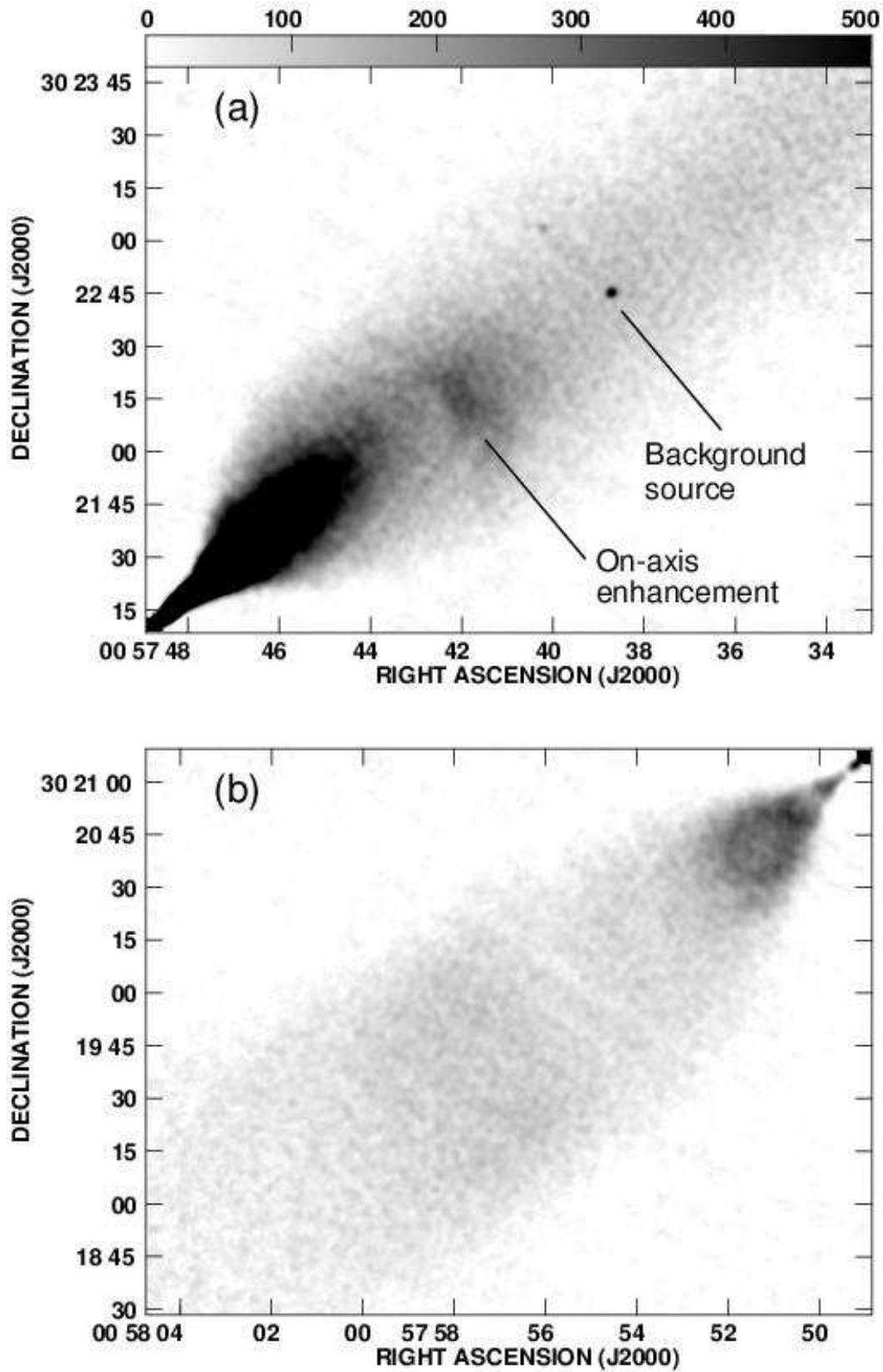}
\caption{Grey-scale of total intensity at 5\,GHz
   corrected for attenuation by the primary beam. The resolution is 2.35\,arcsec
   FWHM and the  grey-scale range is  0 -- 500\,$\mu$Jy (beam area)$^{-1}$ as
   indicated by the labelled wedge. (a) main jet, (b) counter-jet. 
\label{fig:i2.35}}
\end{figure*}

A grey-scale of the large-scale radio structure of NGC\,315 at 327\,MHz
\citep{Mack97} is given in Fig.~\ref{fig:ilow}(a).  Our observations are
sensitive only to emission from the region indicated by the box in this figure
and our mean L-band image of this region at a resolution of 45\,arcsec FWHM is
displayed in Fig.~\ref{fig:ilow}(b). A higher-resolution (5.5\,arcsec FWHM)
image of the same area is shown in Fig.~\ref{fig:i5.5full} and a detail of the
sharp bend in the main jet $\approx$20\,arcmin from the nucleus is plotted at
the same resolution but on a larger scale in Fig.~\ref{fig:ibend}. Finally, the
5-GHz emission from the inner 4\,arcmin of the jets is shown in
Fig.~\ref{fig:i2.35} at a resolution of 2.35\,arcsec FWHM.

We refer to the NW and SE jets as the {\em main} and {\em counter-}jets, as the
former is brighter at most distances from the nucleus.  A striking feature of
the main jet is its almost constant width between $\approx$100 and
$\approx$400~arcsec from the nucleus (Fig.~\ref{fig:i5.5full}). This is the {\em
collimation shoulder} identified by \citet{Willis81}; a similar feature is
visible in the counter-jet, but cannot be traced out as far. The lack of
expansion over such an extended region is surprising if the jets are confined
solely by thermal plasma associated with the surrounding galaxy group, as a
significant pressure gradient would be expected on scales of a few hundred
arcsec (we argue in Section~\ref{RMorigin} that the core radius of the
group-scale plasma is $\approx$225~arcsec). An alternative possibility is that
the jets also respond to the ${\bf J \times B}$ forces of their own toroidal
fields on scales $\ga$100\,arcsec. In Section~\ref{Field}, we show that the
observed polarization structure is consistent with a dominant toroidal component
(see also \citealt{CLBC}), but we cannot tell from the high-frequency
synchrotron emission alone whether this component is vector-ordered or has many
reversals (evidence from Faraday rotation is also inconclusive; see
Section~\ref{RMorigin}).  The possibility that both pressure confinement and
magnetic confinement could act together to produce a collimation shoulder was
discussed by \citet{BCH}, but they assumed rather different physical parameters
from those we now consider appropriate for NGC\,315.  Our observations show that
the collimation shoulder in the main jet ends at a bright feature with a sharp
edge on the side towards the nucleus, inclined by $\approx$60$^\circ$ to the jet
axis. At this point (marked ``Deflection'' in Fig.~\ref{fig:i5.5full}), the flow
changes direction by $\approx$8$^\circ$ and re-expands with an opening angle
$\approx$10$^\circ$ (defined in terms of the jet FWHM and the angular distance
from the nucleus; \citealt{Willis81}). At a similar distance, the counter-jet
does not deflect significantly, its surface brightness decreases monotonically
away from the nucleus and it expands less rapidly than the main jet
\citep[Fig.~\ref{fig:ilow}b]{Willis81}.

The brightness distributions in both jets show large-scale ``banding'' --
repeated, but irregular alternation of bright and faint regions with surface
brightnesses differing by factors of 1.5 to 2 -- along their lengths on
arcminute scales (Fig.~\ref{fig:i5.5full}). The brightness bands extend across
both jets but their variations are slower than those in the flaring region or at
the edges of the jets. These variations could, in principle, result either from
periods of enhanced activity in the nucleus or from interactions between the
jets and their surroundings.  If they were due to fluctuations in activity in
the nucleus that propagated outwards at constant velocity $\beta c$, then they
would appear at projected distances $D_{\rm j}$ and $D_{\rm cj}$ in the main and
counter-jets, respectively, where $D_{\rm j}/D_{\rm cj} =
(1+\beta\cos\theta)/(1-\beta\cos\theta)$ and $\theta \approx$ 38$^\circ$
\citep{CLBC}. Any transverse velocity gradients or deceleration will complicate
this expression and the former effect should distort the bands into arcs that
are concave towards the nucleus. We see no obvious relation between the
distances of the bands in the two jets for any plausible value of $\beta$ and no
evidence for systematic concave curvature of the bands beyond the flaring
region.  Furthermore, the most prominent banding appears to be associated with
regions where the jets deflect or change their collimation properties.  It
therefore seems more likely that the banding is associated with ongoing
interactions between the jets and their surroundings, although we cannot rule
out a contribution to large-scale brightness fluctuations from slow variations
in the jet output.

The remarkable 180$^\circ$ bend in the main jet at the West end of the source is
well known from earlier observations. Our data (Fig.~\ref{fig:ilow}b) show the
emission after the bend at a resolution comparable to the 610-MHz WSRT image
presented by \citet{Mack97}. The brightness distribution at the first part of
the bend (where the jet deflects by $\approx$100$^\circ$) shows complex
structure at 5.5-arcsec resolution. A bright ridge runs along the outside edge,
with a lane of reduced emission next to it (both features are labelled on
Fig.~\ref{fig:ibend}). Note that the suggestion that the flow is re-energised by
an intergalactic shock \citep{Ensslin01} applies to emission further downstream,
after the second bend.  The compact knot of emission at the SW edge of the jet
just before the bend (Fig.~\ref{fig:ibend}) coincides in location and shape with
the optical emission from the flattened background galaxy FGC 0110 (z =
0.021965) and appears not to be physically related to NGC\,315.

The source at RA 00 57 38.710, Dec. 30 22 44.99 (J2000; labelled ``Background
source'' in Fig.~\ref{fig:i2.35}a) is unresolved and has a flat spectrum. The
polarized flux density and ${\bf E}$-vector position angle vary smoothly across
this position, consistent with addition of an unpolarized point source to the
jet emission.  There is a faint optical counterpart on the Digital Sky Survey
and an X-ray point source is detected at a consistent position with the {\em
ROSAT} PSPC (\citealt{WB}; fig.~2). The source is likely to be a background
quasar, despite its location on the projected jet axis.  There is a strong,
diffuse component of emission on the axis of the main jet (``On-axis
enhancement'' in Fig.~\ref{fig:i2.35}a) at $\approx$225\,arcsec from the
nucleus, with no obvious counterpart in the counter-jet.

The 2.35-arcsec resolution images illustrate the initial flaring of the jets
(discussed in detail by \citealt{CLBC} and Worrall et al., in preparation)
followed by recollimation to an almost uniform diameter
\citep{Willis81,Brid82}. Although similar behaviour is observed in other sources
\citep{LB02a,CL}, the physical scale on which NGC\,315 flares and recollimates
is unusually large. This is probably a consequence of the low external density
\citep{WBH}, and we will explore this idea quantitatively elsewhere, using the
conservation-law approach developed by \citet{LB02b}.  The jets in NGC\,315 bend
slightly as they recollimate, from a position angle of $-48.5^\circ$ close to
the nucleus to $-52.8^\circ$ at distances $\ga$100\,arcsec from the nucleus.
The outer isophotes of the main and counter-jets are very similar before the
bend, but the counter-jet is slightly wider at larger distances. The main jet is
brighter than the counter-jet {\em on-axis} at all distances from the nucleus,
but the counter-jet is brighter at the {\em edge} between $\approx$100 and
200\,arcsec. In the flaring region, by contrast, the main jet is significantly
brighter than the counter-jet both on-axis and at the edge of the jet
\citep{CLBC}.

We successfully fit the structure of the inner $\pm$70\,arcsec of the jets in
NGC\,315 using an intrinsically symmetrical model in which all apparent
differences between the main and counter-jets result from relativistic
aberration and beaming \citep{CLBC}. Such models clearly cannot be continued to
indefinitely large scales, as FR\,I jets almost always show evidence (e.g.\
bends and disruption) for asymmetric interaction with the external medium. The
largest-scale structure of NGC\,315 shown in Fig.~\ref{fig:ilow}(a) is an
excellent example, with jets of unequal length terminating in entirely different
ways. This raises the question: how large is the region over which symmetrical,
relativistic models can be applied? At 400~arcsec from the nucleus, there are
clear asymmetries in both the deflection and collimation properties; these must
be intrinsic, although the overall brightness asymmetry persists. There is also
good evidence for interaction with the external medium at distances of 70 --
100\,arcsec on both sides of the nucleus, where both jets bend (symmetrically,
this time). It seems likely that intrinsic and relativistic effects become
comparable between 100 and 200~arcsec, where the sidedness difference on-axis is
the same as on small scales, but the edge value is reversed.  Asymmetries in
apparent magnetic-field structure, which also imply that the flow remains mildly
relativistic in this region, are discussed in Section~\ref{Field}.  Our working
hypothesis is therefore that relativistic effects dominate the observed
differences between the two jets only before the first bends, but that
environmental effects become first comparable and then dominant at larger
distances, although slightly relativistic bulk flow probably continues to the
largest scales and may remain responsible for the generally brighter appearance
of the NW jet far from the nucleus.

\section{Spectra}
\label{Spectra}

\subsection{Accuracy}

We define spectral index $\alpha$ in the sense $S(\nu) \propto \nu^{-\alpha}$.
We estimated spectral indices both for individual pixels and by integration of
flux density over well-defined regions.  Values at 5.5-arcsec resolution were
determined from weighted power-law fits to data at all five frequencies between
1.365 and 5\,GHz.  At 1.5-arcsec resolution, spectral indices were calculated
between 1.413 and 5\,GHz.

There are three main sources of error in the estimate of $\alpha$, as
follows.
\begin{enumerate}
\item The transfer of the amplitude scale from the primary calibrator. The
  errors for the four L-band frequencies are likely to be tightly correlated,
  since they were observed during the same periods, data being taken
  simultaneously at 1365 and 1413\,MHz, and at 1485 and 1665\,MHz. Consequently,
  the principal effect of flux-density scale transfer errors is a constant
  offset in spectral index.
\item Residual deconvolution effects, typically on scales of 5 --
  20\,arcsec. These are approximately proportional to surface brightness.
\item Thermal noise.
\end{enumerate}
We model the error from (ii) as 0.03 times the flux density and that from (iii)
as the noise level estimated off-source (from Table~\ref{noise}), appropriately
integrated. These two contributions are added in quadrature. In addition, we
estimate the rms spectral-index offset due to transfer errors in the
flux-density scale to be 0.02. This should be taken in addition to the errors
quoted below.

\subsection{Spectral-index images and tomography}
\label{tomography}

\begin{figure}
\includegraphics[width=8.5cm]{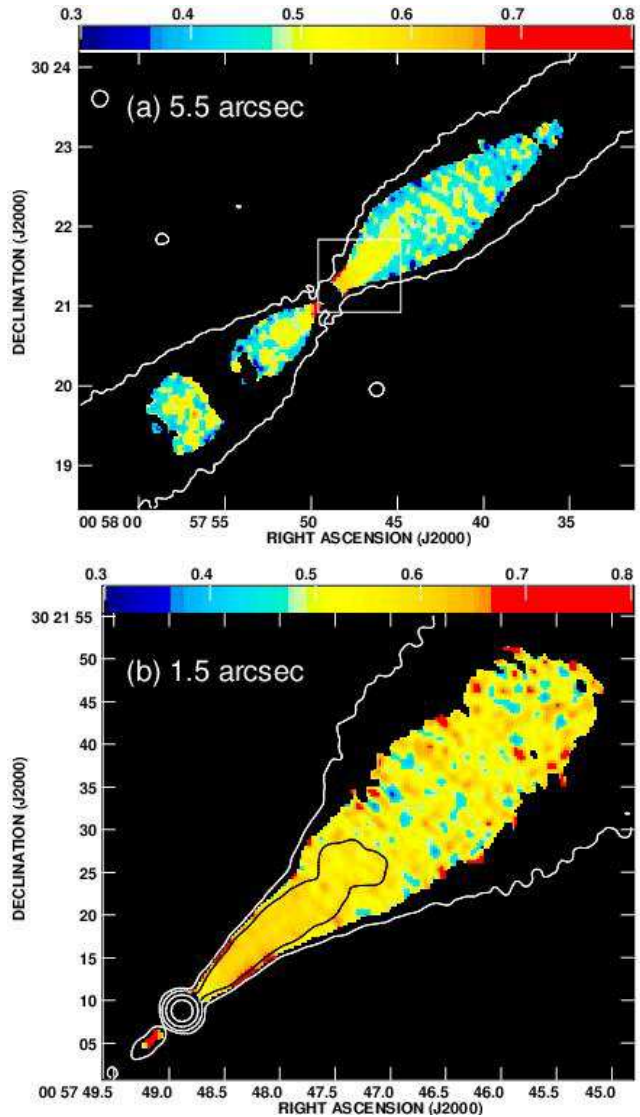}
\caption{(a) False-colour plots of spectral index, $\alpha$ over the range 0.3
  -- 0.8.  (a) $\alpha$ determined by weighted power-law fits to data at all
  five frequencies between 1.365 and 5\,GHz. The resolution is 5.5\,arcsec
  FWHM. Data are plotted only where the rms error in $\alpha$ (excluding
  flux-density scale offsets) is $<$0.05 and the signal-to-noise ratio at all
  frequencies is $>$5. All of the images were made using a zero-spacing flux
  density and the (small) residual zero-point offsets were subtracted before
  fitting. The 75\,$\mu$Jy\,(beam area)$^{-1}$ contour from the mean L-band $I$
  image is also plotted. (b) $\alpha$ between 1.413 and 5\,GHz at a resolution
  of 1.5\,arcsec FWHM for the area shown by the box in panel (a).  Data are
  plotted only where the rms error in $\alpha$ (again excluding flux-density
  scale offsets) is $<$0.05 and the difference between estimates of $\alpha$
  before and after zero-level subtraction $<$0.01. Representative contours of
  total intensity at 5\,GHz are also plotted.
\label{fig:spec}}
\end{figure}

Figs~\ref{fig:spec}(a) and (b) show the spectral indices, $\alpha$, determined
from a weighted power-law fit to data at all five frequencies between 1.365 and
5\,GHz at 5.5\,arcsec resolution and between 1.413 and 5\,GHz at a resolution of
1.5\,arcsec, respectively.

\begin{figure}
\includegraphics[width=8.5cm]{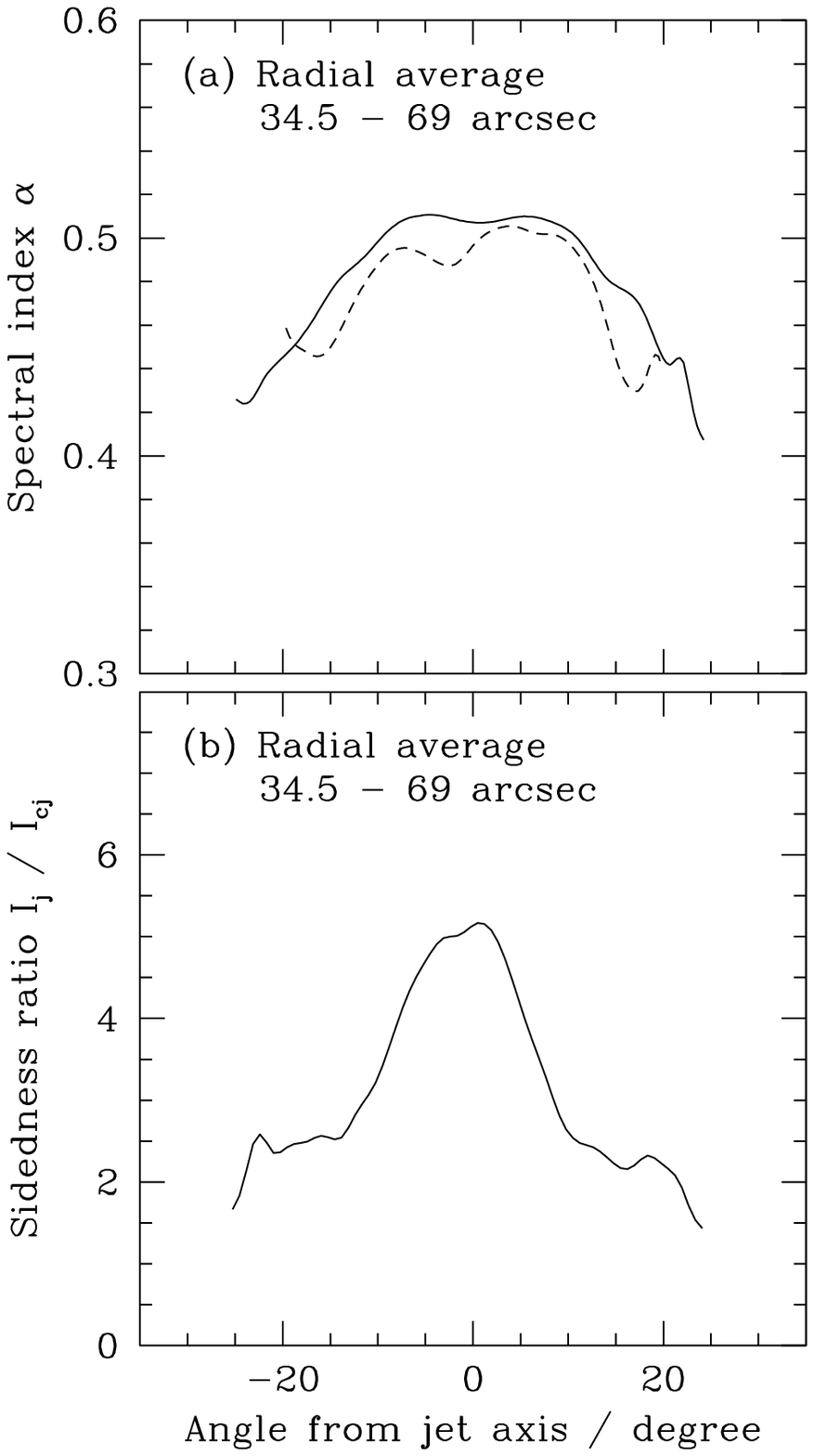}
\caption{(a) Averaged transverse profile of the spectral index determined from a
  weighted power-law fit to data at all five frequencies between 1.365 and
  5\,GHz, as in Fig.~\ref{fig:spec}(a). The resolution is 5.5\,arcsec FWHM. The
  data were averaged along radii from the nucleus between 34.5 and 69\,arcsec
  and are plotted against angle from the jet axis, here taken to be in position
  angle $-48.5^\circ$. Full curve: main jet, dashed curve: counter-jet. (b)
  Profile of jet/counter-jet sidedness ratio, from the data in \citet{CLBC},
  averaged as in panel (a), but with a resolution of 2.35\,arcsec FWHM.
\label{fig:transspec_radial}}
\end{figure}

\begin{figure}
\includegraphics[width=8.5cm]{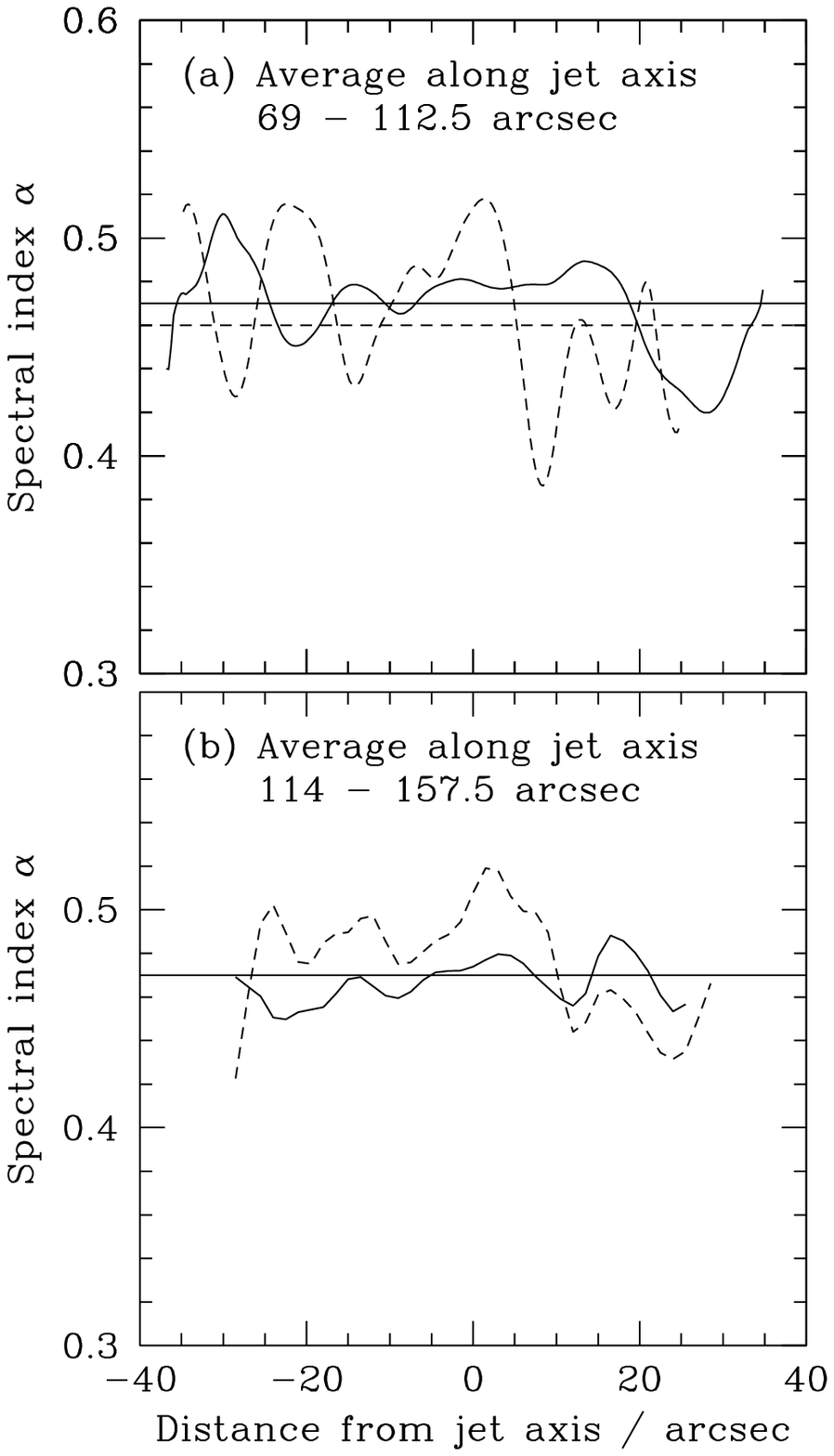}
\caption{Averaged transverse profiles of the spectral index determined from
  weighted power-law fits to data at all five frequencies between 1.365 and
  5\,GHz, as in Fig.~\ref{fig:spec}(a). The resolution is 5.5\,arcsec FWHM. The
  data were averaged along the local jet axis, taken to be in position angle
  $-52.8^\circ$. Full curve: main jet; dashed curve: counter-jet. The horizontal
  lines indicate the mean spectral indices for the regions, which are close to
  0.47 in all cases. (a) 69 -- 112.5\,arcsec, determined by averaging the
  spectral-index image. (b) 114 -- 157.5\,arcsec, determined by summing
  individual $I$ images and making weighted power-law fits (the spectral-index
  image is too noisy to be averaged towards the jet edges).
\label{fig:transspec_long}}
\end{figure}

It is clear from Fig.~\ref{fig:spec}(a) that there are transverse gradients in
spectral index where the jets are expanding rapidly. These gradients can only be
see clearly on spectral-index images where the errors are small, and for this
reason both panels of Fig.~\ref{fig:spec} are blanked where the rms error in
$\alpha$ is $>0.05$.  In order to search for transverse variations over a larger
area, we need to average along the jets. The gradients between 34.5 and
69\,arcsec from the nucleus are best displayed by averaging along radii from the
nucleus and plotting the results against angle from the local jet axis, as is
shown for the main and counter-jets in
Fig.~\ref{fig:transspec_radial}(a). Further from the nucleus, where the jets
recollimate, we have averaged along the local jet axis to derive transverse
spectral-index profiles. The results are shown for two regions in each of the
main and counter-jets in Fig.~\ref{fig:transspec_long}. The fluctuations in
these regions are dominated by quasi-periodic deconvolution errors: this problem
is particularly acute for the counter-jet at $\sim$100\,arcsec from the nucleus
(Fig.~\ref{fig:transspec_long}a). The spectral index is everywhere consistent
with the mean value of $\langle \alpha \rangle = 0.47$ between 70 and
160\,arcsec.

\begin{figure}
\includegraphics[width=8.5cm]{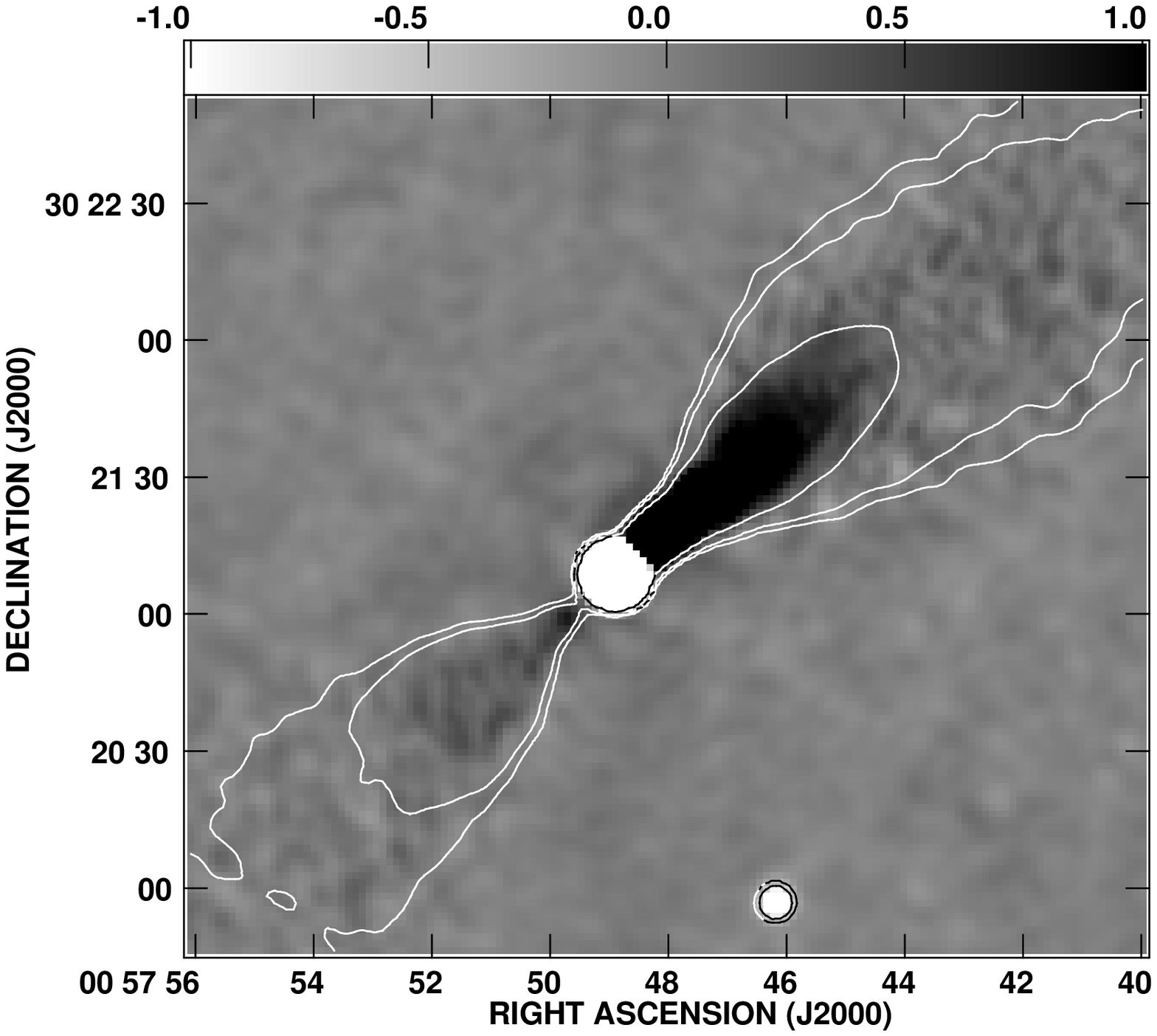}
\caption{The grey-scale plot shows a difference image $I(\nu_1) -
  (\nu_2/\nu_1)^{\alpha_{\rm t}} I(\nu_2)$ for $\nu_1 = 1.413$\,GHz, $\nu_2 =
  5$\,GHz and $\alpha_{\rm t} = 0.44$ over the range $-1$ to +1\,mJy\,(beam
  area)$^{-1}$. The resolution is 5.5\,arcsec FWHM. A few contours of the 5\,GHz
  $I$ image are superposed to outline the jet structure.
\label{fig:spec_tomo}}
\end{figure}

Another method of displaying spatial variations of the spectrum is ``spectral
tomography'' \citep{K-SR,KSetal}. This involves the generation of a set of
images $I_{\rm t}({\bf r}) = I({\bf r},\nu_1) - (\nu_2/\nu_1)^{\alpha_{\rm t}}
I({\bf r},\nu_2)$ for a range of values of $\alpha_{\rm t}$, where $I({\bf
r},\nu)$ is the brightness at position ${\bf r}$ and frequency $\nu$.  If the
brightness distribution can be represented as the sum of two components with
different spectral indices $I({\bf r},\nu) = S_{\rm a}({\bf r})\nu^{-\alpha_{\rm
a}} + S_{\rm b}({\bf r})\nu^{-\alpha_{\rm b}}$, then the ``a'' component will
disappear from the image $I_{\rm t}$ when $\alpha_{\rm t} = \alpha_{\rm a}$. We
made a set of images of $I_{\rm t}$ with $\nu_1 = 1.365$\,GHz, $\nu_2 = 5$\,GHz
and $\alpha_{\rm t}$ from 0.4 -- 0.7 in steps of 0.01.  For $\alpha_{\rm t}
\approx 0.44$, the outer edges of both jets disappear at distances from the
nucleus between $\approx$22\,arcsec and $\approx$80\,arcsec. The image of
$I_{\rm t}$ for $\alpha_{\rm t} = 0.44$ is shown in
Fig.~\ref{fig:spec_tomo}.   Further from the nucleus, $I_{\rm t}$ is close to
zero across the whole width of both jets for $\alpha_{\rm t} \approx
0.47$. There is no single value of $\alpha_{\rm t}$ for which the steep-spectrum
component vanishes completely in an image of $I_{\rm t}$.

The main features of the spectral-index distribution are as follows.
\begin{enumerate}
\item Variations in the spectral index are subtle ($0.4 \la \alpha \la
  0.65$ everywhere).
\item The spectral index is slightly, but significantly steeper in the jet bases
  than elsewhere. Between 7.5 and 22.5\,arcsec from the nucleus the mean values
  at 5.5-arcsec resolution are 0.63 and 0.58 in the main and counter-jets,
  respectively (the difference between them is not significant).
\item At 1.5 arcsec resolution (Fig.~\ref{fig:spec}b), the spectral index of the
  main jet is essentially constant for the first $\approx$15\,arcsec, with a
  mean spectral index $\langle\alpha\rangle = 0.61$, consistent with the value
  determined at lower resolution. 
\item Between $\approx$15 and 70 arcsec from the nucleus, the spectral index is
  steeper on-axis ($\alpha \approx 0.5$) than at the edges ($\alpha \approx
  0.44$) in both jets. This is illustrated by the transverse profiles averaged
  between 34.5 and 69\,arcsec from the nucleus 
  (Fig.~\ref{fig:transspec_radial}a).
\item At 1.5-arcsec resolution, the on-axis spectral index is slightly higher
  between 15 and 60\,arcsec from the nucleus ($\langle\alpha\rangle = 0.55$;
  Fig.~\ref{fig:spec}b) than at smaller distances. $\alpha$ cannot be determined
  to adequate accuracy at the edges of the jet for distances $\ga$15\,arcsec at
  this resolution.
\item The tomographic analysis shows the spectral gradient in a different way:
  if we subtract off a component with $\alpha_{\rm t} = 0.44$
  (Fig.~\ref{fig:spec_tomo}), the emission at the edges of the jet and at large
  distances from the nucleus essentially vanishes. What remains (positive in
  Fig.~\ref{fig:spec_tomo}) corresponds to the jet bases and to a ridge of
  steeper-spectrum emission at larger distances. The latter is clearly visible
  in both jets.
\item The flatter-spectrum edge first becomes detectable at $\approx$15\,arcsec
from the nucleus and widens thereafter. It occupies the entire width of the jet
from $\approx$70\,arcsec outwards. The transition between steeper
and flatter spectrum on-axis is poorly defined.
\item There is no evidence for any transverse spectral gradient at larger
  distances, after the jets recollimate, although the data are noisy and do not
  cover quite the full width of the jets (Fig.~\ref{fig:transspec_long}).
\item This is confirmed by the tomographic analysis: $I_{\rm t}$ for the outer parts
  of the region vanishes for $\alpha_{\rm t} = 0.47$, the mean spectral index,
  confirming that $\alpha$ is constant within our errors.
\end{enumerate}

\subsection{Deprojection of the spectral-index distribution}

\begin{figure}
\includegraphics[width=8.5cm]{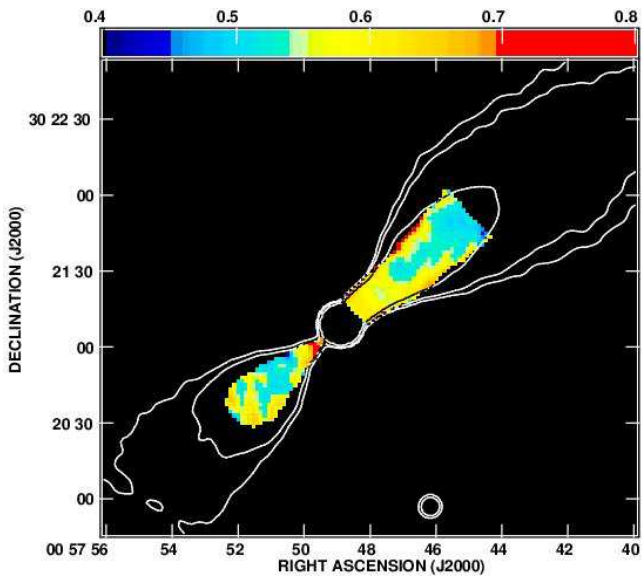}
\caption{The false-colour plot shows an estimate of the spectral index for the on-axis
  emission component alone in the range $0.4 \leq
  \alpha \leq 0.8$. The spectral-index distribution is truncated at a distance
  of 66.5\,arcsec along the axis and is shown only where the residual surface
  brightnesses exceed 0.5 and 1\,mJy\,(beam area)$^{-1}$ at 5 and 1.365\,GHz,
  respectively. A few contours of the 5\,GHz $I$ image are superposed to outline
  the jet structure. The area covered is the same as that in
  Fig.~\ref{fig:spec_tomo} and the resolution is 5.5\,arcsec FWHM.
\label{fig:spinespec}}
\end{figure}

\citet{KSetal} and \citet{K-SR} suggested that the spectral index of an {\em
on-axis} component in a jet is the value of $\alpha_{\rm t}$ at which the
component appears to vanish against the background of the surrounding emission
(exactly as for an {\em edge} component such as that in the NGC\,315 jets;
Section~\ref{tomography}) and can therefore be derived simply from a tomographic
analysis. This requires an additional assumption which may not be correct,
namely that when the the on-axis component is subtracted, the remaining emission
has a smooth brightness distribution (this is {\em not} true for the model
proposed below).  For NGC\,315, our three-dimensional model of the emissivity
\citep{CLBC} gives a good fit to the observed emission at 5\,GHz, so we isolated
the on-axis component and measured its spectral index, as follows.
\begin{enumerate}
\item We first used the tomography image with $\alpha_{\rm t} = 0.44$ as a template for
  the on-axis component. To a reasonable approximation, this defines a cone with a
  half-opening angle of 13.6$^\circ$ projected on the sky. Assuming an angle to
  the line of sight of $\theta = 38^\circ$ \citep{CLBC}, 
  the half-opening angle in the jet frame is $8.3^\circ$ .
\item We then made a model 5-GHz image, as in \citet{CLBC} but with the
  emissivity set to zero within this cone, convolved it to a resolution of 5.5
  arcsec and subtracted it from the observed 5-GHz $I$ image. The residual
  corresponds to the observed emission from within the central cone.
\item We scaled the model to 1.365\,GHz assuming a spectral index of 0.44, as
  appropriate for the edge emission and then subtracted it from the observed
  1.365-GHz image.
\item Both residual images showed little emission towards the edges of the jets,
  implying that the model subtraction was reasonably accurate.
\item Finally, we derived a spectral-index image for the on-axis component
  alone. This is shown in Fig.~\ref{fig:spinespec}.
\end{enumerate} 

\begin{figure}
\includegraphics[width=8.5cm]{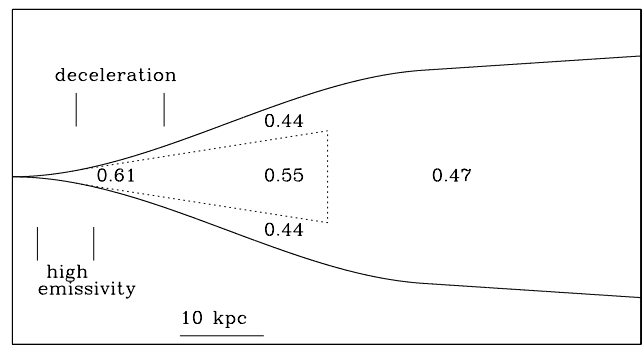}
\caption{A sketch of our proposed three-dimensional distribution of spectral
  index in the jets of NGC\,315. The sketch is in a plane containing the jet
  axis, assuming an angle to the line of sight of 37.9$^\circ$ \citep{CLBC} and
  a linear scale is given. 10\,kpc along the jet projects to 18\,arcsec on the
  plane of the sky. The values of $\alpha$ at various points in the jet are
  marked, together with the approximate extent of the steep-spectrum component
  (dotted) and the regions of high emissivity and rapid deceleration
  \citep{CLBC}.
\label{fig:specsketch}}
\end{figure}

Fig.~\ref{fig:spinespec} shows that the spectrum of the on-axis component
flattens slightly with distance from the nucleus. The mean values of $\alpha$
between 7.5 and 22.5\,arcsec are 0.60 and 0.61 for the main and counter-jets,
respectively. The corresponding values for distances between 22.5 and 66.5
arcsec are 0.56 and 0.55.  For the main jet, these values are consistent with
the measurements at 1.5-arcsec resolution without subtraction. At 5.5-arcsec
resolution in both jets, they are slightly higher than the values measured
on-axis before subtraction, which include a contribution from the
flatter-spectrum, off-axis component.  We conclude that the structure observed
in Fig.~\ref{fig:spec}(a) does not result entirely from the superposition of two
components with constant, but different spectral indices.  The sketch in
Fig.~\ref{fig:specsketch} summarizes our results on the distribution of spectral
index in the jets.

\subsection{Comparison with other sources}

Typical spectral indices for the flaring regions of other FR\,I radio jets are
in the range 0.5 -- 0.6 (\citealt{Young} and references therein). Indeed,
\citet{Young} suggested that FR\,I jets have a ``canonical'' low-frequency
spectral index of $\alpha = 0.55$, but this conclusion is based primarily on
lower-resolution data than we consider here. Their canonical value is
intermediate between that for the jet bases in NGC\,315 and the significantly
flatter spectra seen at larger distances.

There are very few published studies of spectral {\em variations} in the flaring
regions of FR\,I jets.  Short regions of slightly steeper spectrum than the
average have been detected in the bases of three other FR\,I jets: 3C\,449
\citep{K-SR}, PKS\,1333$-$33 \citep{KBE} and 3C\,31 (Laing et al., in
preparation).  The measurement of $\alpha = 0.7 \pm 0.2$ in 3C\,449 applies to
the faint inner jets (corresponding to the innermost 5\,arcsec in NGC\,315), but
differs marginally from that of the brighter nearby emission, given the large
error. In PKS\,1333$-$33, the spectrum flattens from $\alpha \approx 1.0$ to
$\alpha \approx 0.6$ between 10 and 35\,arcsec from the nucleus (2.4 -- 8.4\,kpc
in projection); this includes both the faint inner jets and the bright part of
the flaring region, as in NGC\,315.  In 3C\,31, there is a steeper-spectrum
region extending to $\approx$6\,arcsec in the main jet (plausibly also in the
counter-jet).

There is a flatter-spectrum edge on one side of the main jet in 3C\,31 (Laing et
al., in preparation). Transverse variations of spectral index in the sense that
the spectrum is {\em steeper} on-axis have not been reported in any other
sources, but their flaring regions are smaller in both linear and angular size
and are difficult to resolve. Gradients at the level of $\Delta\alpha \approx$
0.05 -- 0.1 are also tricky to detect without data at more than two
frequencies.

The tendency for the jet spectrum to be flatter at the edge is in the opposite
sense to the spectral gradients found on large scales in tailed radio sources
\citep[Laing et al., in preparation]{KSetal,K-SR} but the latter effect occurs
in completely different regions of the jets, where they merge into the tails.

\subsection{Acceleration mechanisms}

The X-ray emission detected by \citet{WBH} in NGC\,315 coincides with the
steeper-spectrum ($\alpha =0.61$) region at the base of the main jet. The form
of the synchrotron spectrum in those FR\,I jet bases which emit X-rays is now
well established \citep{Hard01,Hard02,Parma03,Hard05,PW05}. It can be
characterised as a broken power law with spectral indices of 0.5 -- 0.6 at radio
wavelengths and 1.2 -- 1.6 in X-rays. The spectrum of NGC\,315 is consistent
with this pattern, although its high-frequency slope is poorly constrained
\citep{WBH}. This spectral shape is consistent with synchrotron emission from a
single electron population; ongoing particle acceleration is therefore required
(see also Worrall et al., in preparation).  The magnetic-field strengths
calculated for the on-axis emissivity model of \citet{CLBC}, assuming a
minimum-pressure condition, range from 3.3\,nT at 6\,arcsec in projection to
0.44\,nT at 69\,arcsec. For electrons with Lorentz factor $\gamma$ radiating at
the synchrotron critical frequency $\nu_{\rm c}$ in a magnetic field $B$, we
have $\nu_{\rm c}$/Hz = 41.99 $\gamma^2$ ($B$/nT) \citep{Longair}, so at 5\,GHz, $6
\times 10^3 \la \gamma \la 1.6 \times 10^4$ and in the X-ray band at 1\,keV, $4
\times 10^7 \la \gamma \la 1.1 \times 10^8$.

The flatter-spectrum edges occur where \citet{CLBC} infer substantial velocity
shear across the jets.  If the jets are relativistic and faster on-axis than at
their edges, the approaching jet always appears more centre-brightened than the
receding one.  This difference is reflected in the sidedness-ratio image. The
average transverse sidedness profile between 34.5 and 69\,arcsec from the
nucleus is shown in Fig.~\ref{fig:transspec_radial}(b) for comparison with the
spectral-index profile over the same region. The velocity profile is modelled as
a truncated Gaussian function with $\beta = 0.38$ on-axis and 0.22 at the edge,
although \citet{CLBC} note that the on-axis velocity may be larger ($\beta
\approx$ 0.5) if the shear occurs over a narrow range of radii in the jet: this
might give a better fit to the sidedness profile.

The flatter-spectrum edge first becomes visible $\approx$15\,arcsec from the
nucleus.  This coincides to within the errors with:
\begin{enumerate}
\item the {\em start} of rapid deceleration, as inferred by \citet{CLBC}, 14\,arcsec in
projection from the nucleus;
\item the {\em end} of the region of enhanced radio and (in the main jet only) X-ray
emissivity \citep[Worrall et al., in preparation]{CLBC}.
\item The first point at which the observed jet/counter-jet sidedness image
  gives any evidence for transverse velocity gradients, 16\,arcsec from the
  nucleus \citep{CLBC}.
\end{enumerate}
The deceleration and enhanced emission regions are marked on
Fig.~\ref{fig:specsketch}. Note that the detection of transverse gradients in
sidedness and spectral index may be limited by resolution.

The changes of spectral index observed in NGC\,315 suggest that (at least) two different
electron acceleration mechanisms are required, as follows.
\begin{enumerate}
\item The first mechanism dominates at the base of the flaring region (the
  initial 15\,arcsec in NGC\,315) and may continue at a lower level on the axis
  of the jet to $\approx$70\,arcsec. It generates emission from radio to X-ray
  wavelengths and has a characteristic spectral index $\alpha \approx 0.6$ in
  the former band.  Three pieces of evidence suggest that this mechanism is
  dominant where the jet is fast ($\beta \approx 0.9$). Firstly, its
  characteristic spectral index is observed across the whole of the jet width in
  NGC\,315 until the start of rapid deceleration. Secondly, in both 3C\,31
  \citep{LB04} and NGC\,315 \citep{WBH}, the bright X-ray emission occurs
  upstream of the deceleration region.  Finally, \citet{LB04} show from radio
  data alone that significant injection of fresh relativistic particles is
  required before the start of deceleration in 3C\,31 to counter-balance
  adiabatic losses.
\item The second mechanism causes the flattening of the spectrum towards the
  edges of the jets observed from 15\,arcsec outwards, but eventually spreading
  over the entire jet width. It produces spectral indices in the range $0.44 \la
  \alpha \la 0.5$ for electrons emitting at radio wavelengths in NGC\,315 and
  appears to be associated with velocity shear across the jets.  A possible
  candidate for the this mechanism is the shear acceleration process described
  by \shortcite{RD1,RD2}, but their estimates of the acceleration timescale for
  electrons in conditions appropriate to FR\,I jets are very long (at least if
  the mean free path $\sim$ gyro-radius), and it is unclear whether the process
  is efficient enough to influence the spectrum.
\end{enumerate}

\begin{figure*}
\includegraphics[width=17cm]{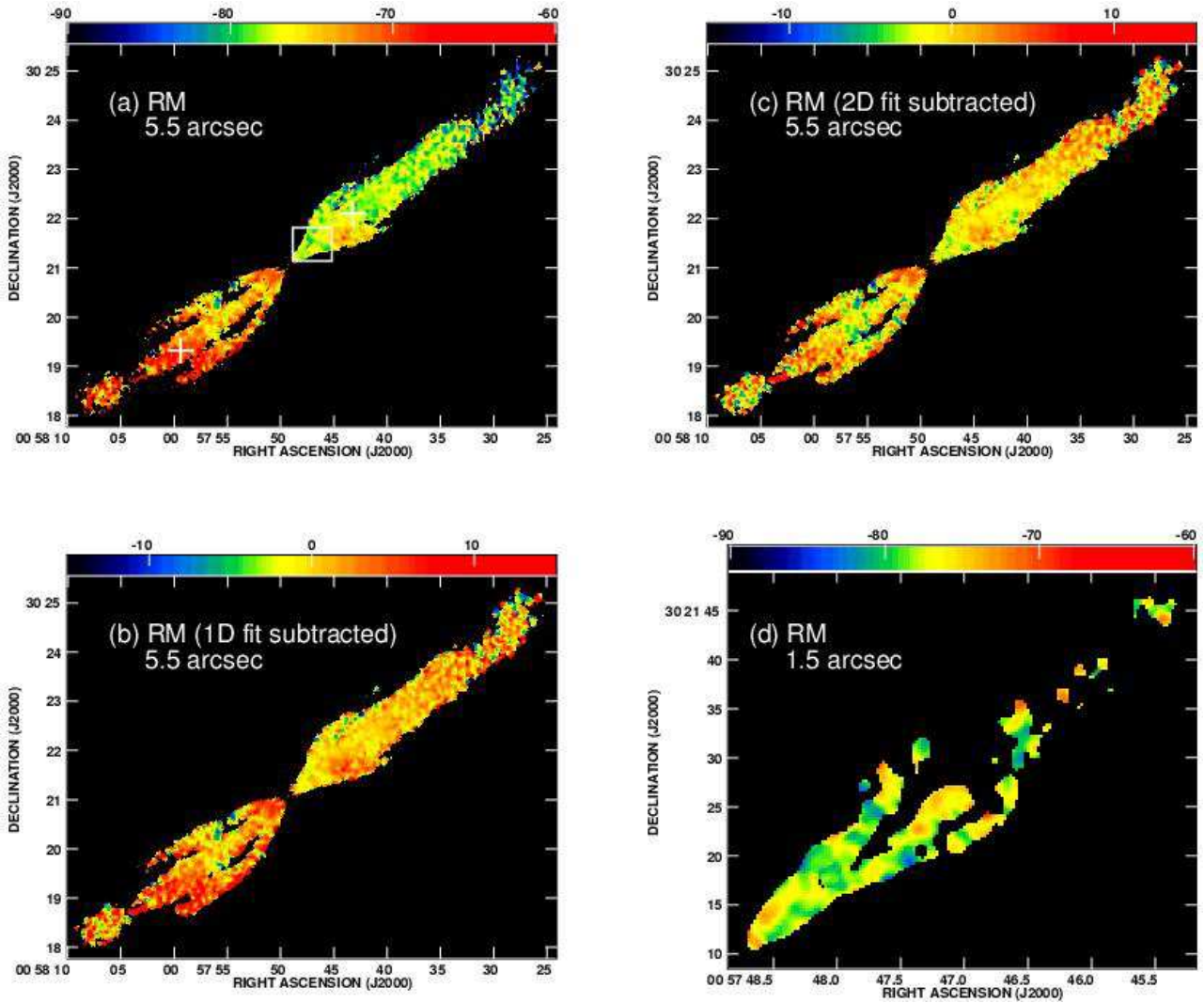}
\caption{False colour images of rotation measure for the jets in NGC\,315. In panels
  (a)-- (c), the RM is derived from fits to {\bf E}-vector position angles at 5
  frequencies.  (a) Observed RM, in the range 
  $-90$ to $-60$ rad\,m$^{-2}$. The crosses mark the positions corresponding to the $\chi$ --
  $\lambda^2$ plots in Fig.~\ref{fig:rmegs}. (b) RM after subtraction of a linear function
  of distance along the jet derived from an unweighted fit to the data in the range $-15$
  to $+15$ rad\,m$^{-2}$. (c) As (b), but for a fit to gradients along and
  across the jets. (d) RM at a resolution of 1.5\,arcsec FWHM, derived
  from the position-angle difference between 1.413 and 5\,GHz, using the
  lower-resolution RM image to resolve $n\pi$ ambiguities. Data are plotted
  only where the rms error in RM $<$2\,rad\,m$^{-2}$. The area covered by this
  figure is indicated by the box on panel (a).
\label{fig:RMimages}}
\end{figure*}

\begin{figure}
\includegraphics[width=8.5cm]{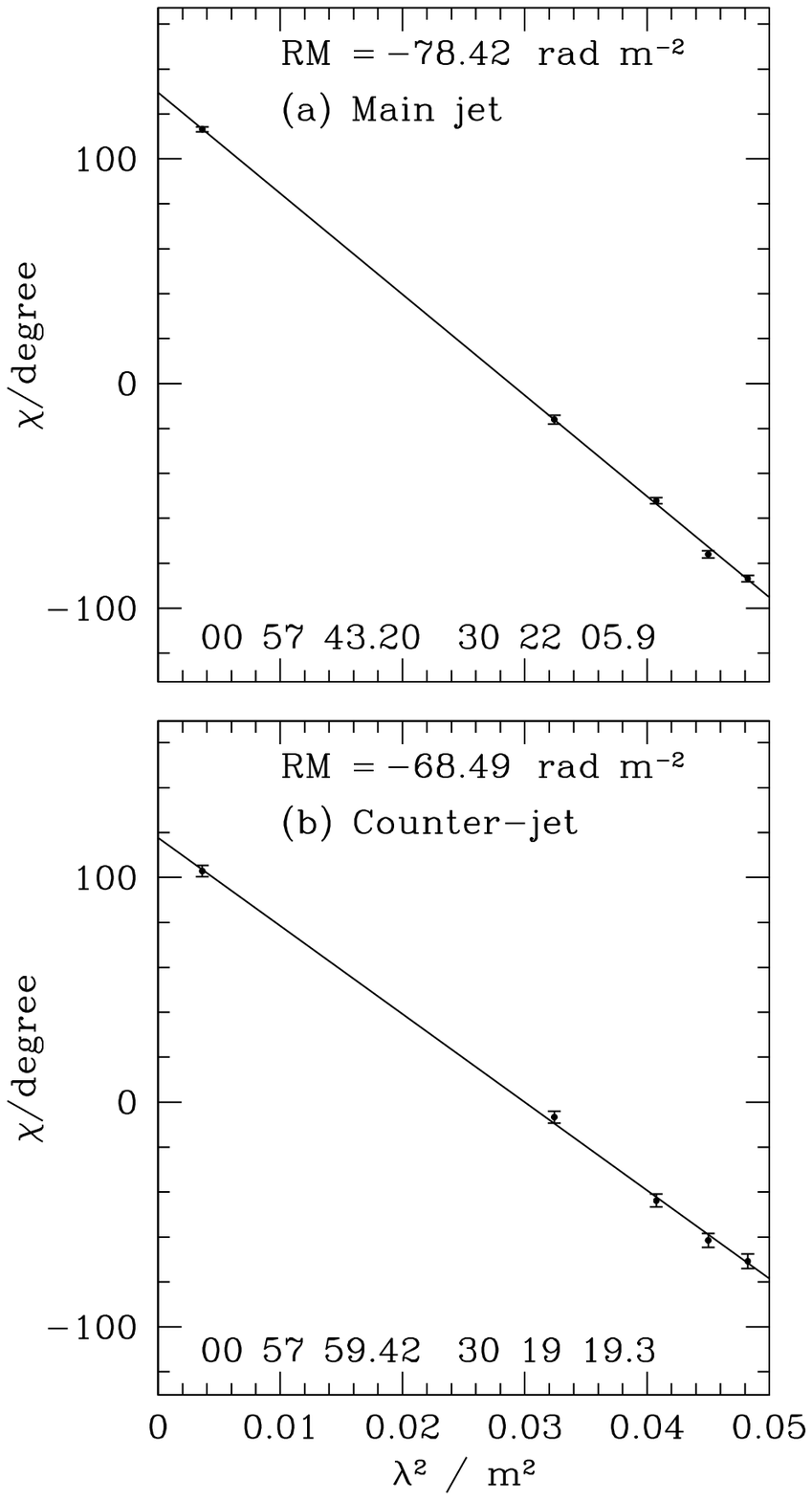}
\caption{Plots of {\bf E}-vector position angle, $\chi$, against $\lambda^2$ for
  representative positions in the main jet (panel a) and counter-jet (panel
  b). The coordinates are given on the plots and the positions are those
  marked by crosses in Fig.~\ref{fig:RMimages}(a).
\label{fig:rmegs}}
\end{figure}

\begin{figure}
\includegraphics[width=7cm]{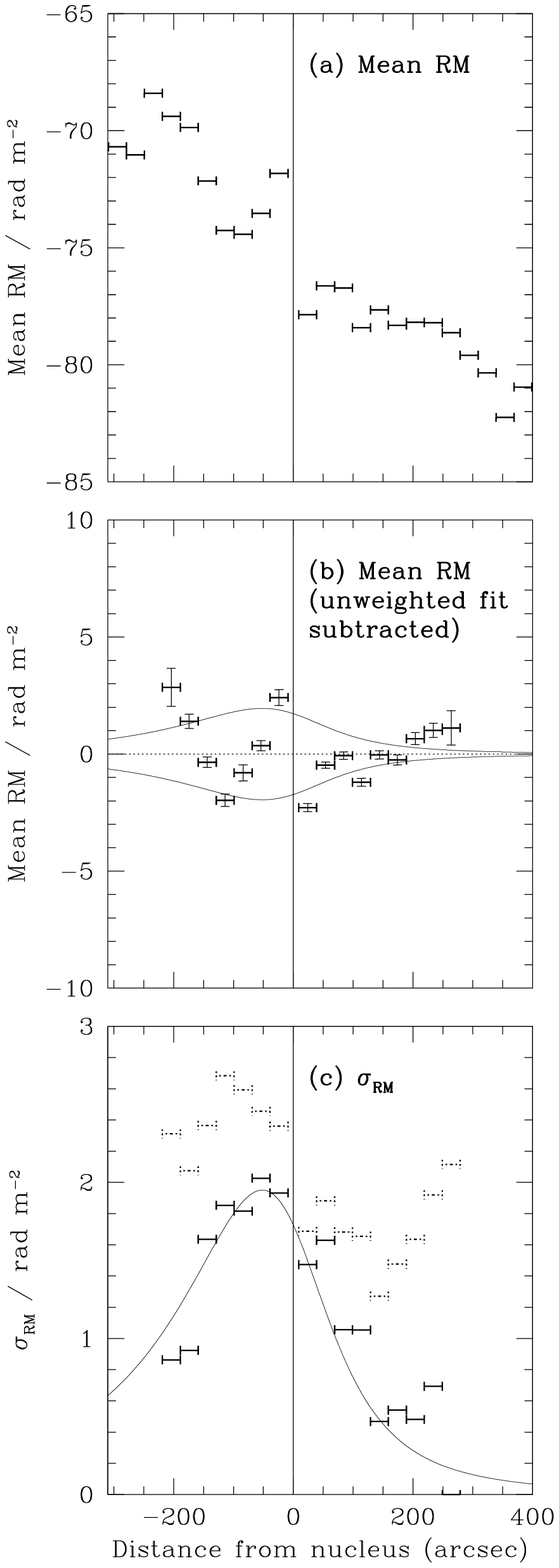}
\caption{Variation of the mean and rms RM along the jets. The average jet axis
  is taken to be at position angle $-52.8^\circ$.  (a) A profile of
  RM averaged over boxes of length 30\,arcsec along the jet axis.  (b) Mean RM
  profile, after subtraction of a linear fit. Only points with fitting error
  $\leq$2.5\,rad\,m$^{-2}$ were used and boxes are only plotted if they contain
  more than 50 such points. The vertical bars indicate the error on the
  mean.  (c) the rms RM, with respect to the
  mean of the points in the box. Dashed bars denote the uncorrected values,
  $\sigma_{\rm RM raw}$; full bars show the values after making a first-order
  correction for fitting error, $\sigma_{\rm RM}$ (see text). The curve
  plotted in panels (b) and (c) is the model rms RM profile described in the text.
  \label{fig:RMprofiles}}
\end{figure}

\begin{figure}
\includegraphics[width=7cm]{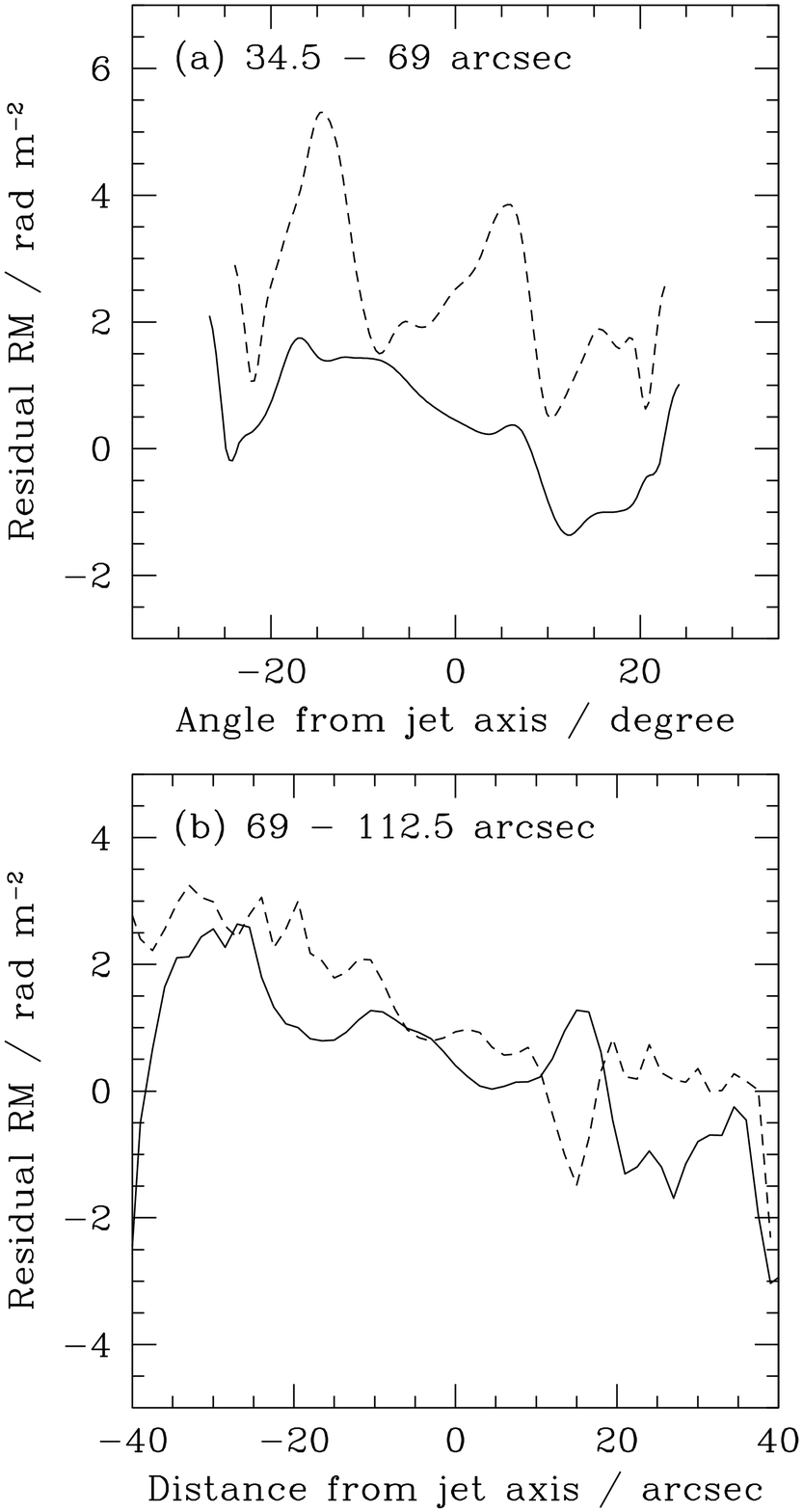}
\caption{Transverse variation of RM after subtracting a uniform
  gradient along the axis for the main jet (full line) and counter-jet (dashes). (a)
  Averages along radii from the nucleus between 34.5 and 69\,arcsec plotted
  against the angle from the axis, which is taken to be in position angle
  $-48.5^\circ$. (b) Averages along the jet (in position angle $-52.8^\circ$)
  between 69 and 112.5\, arcsec, plotted against distance from the axis.
  \label{fig:RMtrans}}
\end{figure}

\section{Faraday rotation and depolarization}
\label{Faraday}

\subsection{Faraday rotation}
\label{RM}

In order to investigate the variations of Faraday rotation along the jets of
NGC\,315, we made images of rotation measure (RM) at a resolution of 5.5\,arcsec
by least-squares fitting to the relation $\chi(\lambda^2) = \chi(0) + {\rm
RM}\lambda^2$ (where $\chi$ is the {\bf E}-vector position angle) for all 5
frequencies between 1.365 and 5\,GHz. The fits were weighted by errors in $\chi$
derived from Table~\ref{noise}. We excluded a small region around the core which
was affected by residual instrumental polarization and included only points
where the rms error in position angle was $<$15$^\circ$ at all frequencies. The
resulting RM image is shown in Fig.~\ref{fig:RMimages}(a). The fit to a
$\lambda^2$ law is very good everywhere: two examples are shown in
Fig.~\ref{fig:rmegs}. The extreme value of RM $\approx -90$\,rad\,m$^{-2}$
results in rotations of $\approx1^\circ$ between the two centre frequencies
combined in our 5\,GHz dataset and $\approx$0.4$^\circ$ and $\approx$5$^\circ$
across the bands at 5 and 1.4\,GHz, respectively. The worst of these effects,
rotation across the band at the lowest frequencies, results in a spurious
depolarization $<$0.1\%, which is negligible compared with errors due
to noise. The images of $Q$ and $U$ show little power on large spatial scales
and our estimates of position angle should be reliable over the full range of
distances from the nucleus shown in Fig.~\ref{fig:RMimages}(a). The area over
which the RM can be determined accurately is limited primarily by the primary
beam at 5\,GHz (540\,arcsec FWHM).

The mean RM is $-$75.7\,rad\,m$^{-2}$. The most obvious feature of the RM image
is a nearly linear gradient along the jets, as shown by the profile in
Fig.~\ref{fig:RMprofiles}(a). In order to reveal smaller-scale structure in the
RM, we initially fitted and subtracted a function ${\rm RM} = {\rm RM}_0 + a_x
x$, where $x$ is measured along the axis from the nucleus ($a_x$ and ${\rm
RM}_0$ are constants). We used an unweighted least-squares fit, as any attempt
to weight by the estimated errors caused the brightest part of the main jet to
be fitted at the expense of other regions. The gradient is $a_x = $
0.025\,rad\,m$^{-2}$\,arcsec$^{-1}$ and the residual image is shown in
Fig.~\ref{fig:RMprofiles}(b). This indicates that variations in RM across the
jet are also significant, as can be seen more clearly in averaged profiles,
particularly between 69 and 112.5\,arcsec from the nucleus
(Fig.~\ref{fig:RMtrans}b). Closer to the nucleus, the gradient is barely visible
(Fig.~\ref{fig:RMtrans}a), but the jets are narrower there and the profiles are
consistent with the gradient measured at larger distances. The transverse
variation also appears to be linear, so we fitted and subtracted a function
${\rm RM} = {\rm RM}_0 + a_x x + a_y y$, where $y$ is a coordinate transverse to
the jet and $a_y$ is a constant. The gradients along and transverse to the jet
axis become $a_x$ = 0.018\,rad\,m$^{-2}$\,arcsec$^{-1}$ and $a_y$ =
0.051\,rad\,m$^{-2}$\,arcsec$^{-1}$, respectively.  Taken at face value, the
best-fitting gradient is 0.054\,rad\,m$^{-2}$\,arcsec$^{-1}$ at an angle of
72$^\circ$ to the mean jet axis. Note, however, that the transverse gradient is
essentially determined by a subset of the data from the widest parts of the main
and counter-jets (Fig.~\ref{fig:RMtrans}b).

Removal of the large-scale gradient leaves fluctuations in the local mean RM
which appear significantly larger on the counter-jet side
(Fig.~\ref{fig:RMimages}c). By definition, the signal-to-noise ratio in $I$ is
lower in the counter-jet.  Although this is partially offset by a higher average
degree of polarization, the errors in RM are still larger than in the main
jet. To evaluate the significance of the fluctuations, we considered only points
with fitting errors $\leq$2.5\,rad\,m$^{-2}$ and calculated the expected errors
in the means for boxes of length 30\,arcsec along the jet axis containing more
than 50 such points (the errors are corrected for oversampling).  The results
are shown in Fig.~\ref{fig:RMprofiles}(b). The fluctuations are significant, and
form an ordered pattern with a typical scale $\sim$100\,arcsec. They are larger
by a factor of $\approx$2 in the counter-jet and the first bin of the main jet
(within 30 arcsec of the nucleus) compared with the rest of the main jet.

The residual fluctuations within the boxes (i.e.\ after subtracting the local
mean) are comparable with the errors in RM except in the brightest regions close
to the nucleus. We therefore made a first-order correction to the rms RM,
$\sigma_{\rm RM raw}$, by subtracting the fitting error $\sigma_{\rm fit}$ in
quadrature to give $\sigma_{\rm RM} = (\sigma_{\rm RM raw}^2 - \sigma_{\rm
fit}^2)^{1/2}$. The profiles of $\sigma_{\rm RM raw}$ and $\sigma_{\rm RM}$ are
both plotted in Fig.~\ref{fig:RMprofiles}(c), for the same selection of points
as in Fig.~\ref{fig:RMprofiles}(b).  The corrected profile is very uncertain,
but suggests that $\sigma_{\rm RM}$ has a maximum of 2\,rad\,m$^{-2}$ close to
the nucleus on the counter-jet side and may be slightly asymmetric in the sense
that the rms RM is lower in the main jet than the counter-jet at the same
distance from the nucleus.

We can image these smaller-scale fluctuations directly only at the bright base
of the main jet. There, the RM at a resolution of 1.5\,arcsec FWHM can be
derived accurately from the difference between 1.413 and 5\,GHz position-angle
images, using the lower-resolution data to resolve the $n\pi$ ambiguity.
Fig.~\ref{fig:RMimages}(d) shows the RM at this resolution. Data are
plotted only where the rms error in the fitted RM
$<$2\,rad\,m$^{-2}$. Fluctuations are clearly detected: the rms is $\sigma_{\rm
RM raw}$ = 2.1\,rad\,m$^{-2}$, giving $\sigma_{\rm RM} =$ 1.6\,rad\,m$^{-2}$
after making a first-order correction for fitting error, as above. This is in
good agreement with the value for the innermost bin of the profile of rms RM for
the main jet at lower resolution (Fig.~\ref{fig:RMprofiles}c).

\subsection{Depolarization}
\label{Depol}

The variation of $p$ with wavelength at low resolution potentially measures
fluctuations of RM across the observing beam which cannot
be imaged directly with adequate sensitivity. This
variation is small, and is best quantified by fitting to the first-order
approximation $p(\lambda^2) \approx p(0) +p^\prime(0)\lambda^2$ where
$p^\prime(\lambda^2) = dp/d(\lambda^2)$. We expect $p^\prime(0) < 0$
(depolarization) under most circumstances.  The quantity $p^\prime(0)/p(0)$ is
directly related to the commonly quoted depolarization ratio, but is biased in
the present case, since both the gradient and the degree of polarization depend
directly on a single high-frequency measurement (at 5\,GHz), so deviations in $p(0)$
and $p^\prime(0)$ are anticorrelated. We therefore tested for the presence of
depolarization using the gradient $p^\prime(0)$ alone.  We derived $p^\prime(0)$
by weighted least-squares fitting to images of $p$ at the 5 frequencies between
1.365 and 5\,GHz at a resolution of 5.5\,arcsec FWHM. Two sets of $p$ images
were used: in the first, an estimate of the local zero-level was subtracted from
the $I$ images before calculating $p = P/I$; in the second, the original $I$
images were used (any differences indicate systematic errors in the estimation
of large-scale structure). The fitting weights were the inverse squares of
errors in $p$ derived from the values in Table~\ref{noise} and points were only
included if the errors were $<$0.3 at all frequencies.  The resulting
polarization gradients are very small, and there are no obvious variations. In
order to determine the significance of the gradients, we measured their mean
values over the main and counter-jets.  The maximum scale of structure imaged
accurately in total intensity is $\approx$300\,arcsec (Table~\ref{Datasets}), so
we calculated the means for points between 9 and 150\,arcsec from the nucleus
along the jet axis (excluding the small region around the core to avoid spurious
instrumental polarization, as in Section~\ref{RM}).  The mean values for the
main and counter-jets derived from the zero-level corrected images were $\langle
p^\prime(0)\rangle = 0.03 \pm 0.08$ and $0.16 \pm 0.16$, respectively. Without
zero-level correction, the values become $0.05 \pm 0.08$ and $0.22 \pm 0.16$.
We conclude that the increase of $p$ with wavelength for $\lambda \leq 0.22$\,m
($\nu \geq 1.365$\,GHz), which is in any case opposite to the expected effects
of Faraday rotation, is not significant.

\subsection{The origin of the rotation measure}
\label{RMorigin}

NGC\,315 has Galactic coordinates $l = 124.^\circ6$, $b = -32.^\circ5$. This is
on the outskirts of Region A of \citet{SK}, where the majority of sources have
large negative RM's ($\sim -100$\,rad\,m$^{-2}$ at the centre of the
region). \citet{DK05} derived spherical-harmonic models of the Galactic RM
distribution by fitting to RM's of large numbers of extragalactic sources. These
models predict a Galactic contribution $\approx
-47$\,rad\,m$^{-2}$ at the position of NGC\,315. The bulk of the mean RM of
$-$75.7\,rad\,m$^{-2}$ is therefore likely to be Galactic in origin.

\citet{SC86} determined RM variations, which they argued to be primarily
Galactic, across a number of sources within Region A (their Region 1). Their
plot of squared RM difference $\Delta{\rm RM}^2$ against separation suggests
$\langle \Delta{\rm RM}^2\rangle^{1/2} \sim$ 10 -- 30\,rad\,m$^{-2}$ on a scale
of 700\,arcsec, but with large uncertainties. The observed $\Delta{\rm RM}$
($\approx 15$\,rad\,m$^{-2}$ on the same scale along the jets of NGC\,315;
Fig.~\ref{fig:RMprofiles}a) is again consistent with a Galactic origin.  The
maximum change in RM across the jet ($\approx 4$\,rad\,m$^{-2}$ over 80\,arcsec;
Fig.~\ref{fig:RMtrans}b) is within the range of the upper limits plotted by
\citet{SC86}.  We note, however, that NGC\,315 is located outside the core of
Region A, so it may be that smaller values of $\langle \Delta{\rm
RM}^2\rangle^{1/2}$ are appropriate.

The linearity of the RM gradient along the jet and the fact that the maximum
gradient is aligned neither with the jet axis nor with the minor axis of the
galaxy both imply that little of the Faraday rotating medium is associated with
the jets or with the host galaxy.  In either case we would expect some
non-linear variation with distance from the nucleus.

We conclude that most of the mean RM and its linear gradient are likely to be
Galactic in origin, but a significant contribution from material local to
NGC\,315 is not ruled out. In particular, we cannot exclude the hypothesis that
some of the RM gradient transverse to the jet results from an ordered toroidal
field within or just outside the jet \citep{L81,Blandford93}. The associated
position angle rotation between 1.365 and 5\,GHz is at most a few degrees and
significant departures from $\lambda^2$ rotation would not be detectable even if
the thermal plasma responsible for the Faraday rotation is mixed with the
synchrotron-emitting material \citep{Burn66}. If local toroidal fields are
solely responsible for the transverse RM gradients, then their vector directions
must be the same in the main and counter-jets.

The {\em residual} RM fluctuations are qualitatively very similar to those in
3C\,31 (Laing et al., in preparation) but have amplitudes that are 10 times
smaller on similar angular scales. The larger-scale ($\sim$100\,arcsec)
fluctuations are systematically lower on the main (approaching) jet side. The
distribution of fluctuations on smaller scales is also consistent with such an
asymmetry, but is not well determined.  As with the transverse gradients
discussed earlier, the observed position-angle rotations are too small to be
sure that the RM fluctuations are due to foreground plasma, but the asymmetry
between approaching and receding jets suggests an origin in a distributed
magnetoionic medium surrounding the host galaxy, a possibility we now explore. 

\begin{figure*}
\includegraphics[width=17cm]{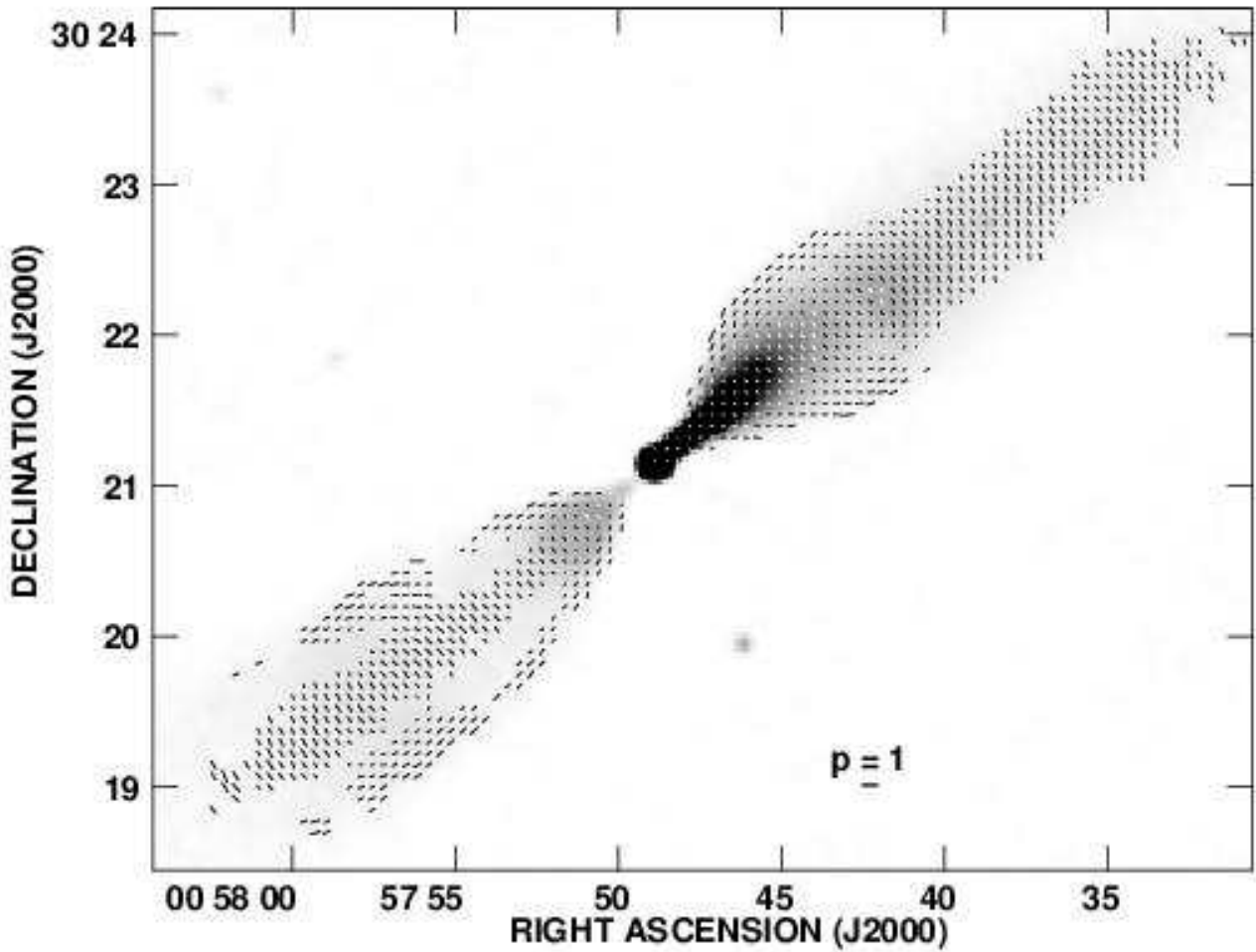}
\caption{Vectors whose lengths are proportional to the degree of polarization at
  zero wavelength, $p(0)$, from the analysis of Section~\ref{Depol} and whose
  directions are those of the apparent magnetic field, derived from the
  5-frequency RM fit of Section~\ref{RM}. The vector scale is indicated by the
  labelled bar. The grey-scale is of total intensity at 4.985\,GHz and the
  resolution is 5.5\,arcsec FWHM.
\label{fig:ivec5.5}}
\end{figure*}

The thermal plasma associated with NGC\,315 and observed using {\em Chandra} can
be described by a beta model with a core radius of 1.55\,arcsec
\citep{WBH}. This cannot be responsible for the RM fluctuations, which occur on
far larger scales. The most likely hot plasma component to be responsible for
the Faraday rotation would be associated with the poor group of galaxies
surrounding NGC\,315 \citep{Nolthenius,Miller02}, but has not yet been detected
in X-ray observations. Since RM fluctuations are seen in both jets, a spherical
distribution is plausible.  In order to make a rough estimate of the parameters
of the putative group component, we took a simple model for the field structure
in which cells of fixed size $l$ at radius $r$ contain randomly orientated
fields $B(r)$ \citep{LD82,Felten96}. The density distribution was taken to be a
beta model: $n(r) = n_0 (1+r^2/r^2_c)^{-3\beta_{\rm atm}/2}$ with $B(r) \propto
n(r)^N$.  We derived the variation of $\sigma^2_{\rm RM}$ along the projection
of the jet axis by numerical integration, assuming that the jets have $\theta =
37.9^\circ$ everywhere. This is a similar approach to the calculation of
depolarization asymmetry by \citet{GC91} and \citet{Tribble92}; our code also
reproduces the analytical results of \citet{Felten96} for $\theta = 90^\circ$
and $N = 0$ or 0.5.  Following \shortcite{Dolag01,Dolag06}, we assumed that
$B(r) \propto n(r)$ or $N = 1$ (an empirical result derived from RM's of radio
sources in and behind cluster and rich groups). Our detection of RM variations
on a range of scales is qualitatively consistent with the idea that the power
spectrum of the magnetic-field fluctuations is a power law
\citep{Tribble91,EV03,Murgia,VE05}, but our data are too noisy and poorly
sampled to constrain its functional form.  \citet{Murgia} show that a
single-scale model gives a very similar relation between the RM variance
$\sigma^2_{\rm RM}$ and radius $r$ to that derived for the more realistic case
of a power-law power spectrum provided that $l$ is interpreted as the
correlation length of the magnetic field. Finally, we fitted the resulting
$\sigma_{\rm RM}$ curves by eye to the profiles of RM fluctuations on different
scales in Fig.~\ref{fig:RMprofiles}(b) and (c).  We fixed the value of
$\beta_{\rm atm} = 0.5$ and adjusted the core radius to give a reasonable fit to
the profile. For both plots, we found that $r_c \approx 225$\,arcsec gave an
adequate fit.  Strictly speaking, $\sigma^2_{\rm RM}$ is the RM variance
evaluated over a window much larger than the maximum fluctuation scale, whereas
the profiles in Fig.~\ref{fig:RMprofiles} describe fluctuations over two
different ranges of scale. We have therefore added the two model variances to
give a rough estimate of the total (the curves shown in
Fig.~\ref{fig:RMprofiles}b and c actually have the same amplitude). Note that
the process of removing a linear trend from the RM profile will have suppressed
some power in large-scale fluctuations, particularly transverse to the jet.  The
amplitude of the model variance profile is related to the central density and
field, and to the correlation length \citep{Felten96}: $(n_0/{\rm m}^{-3})^2
(B_0/{\rm nT})^2 (l/{\rm kpc}) \approx 700$.  This is a very rough estimate, but
is enough to establish that a low-density group-scale gas component with a low
magnetic field can generate the observed Faraday rotation. Plausible parameters
might be $n_0 \approx$ 600\,m$^{-3}$, $B_0 \approx$ 0.015\,nT (0.15\,$\mu$Gauss)
and $l \approx$ 10\,kpc. We note that the recollimation of the jets may also
require a large-scale hot gas component.

\section{Magnetic-field structure}
\label{Field}

\begin{figure*}
\includegraphics[width=15cm]{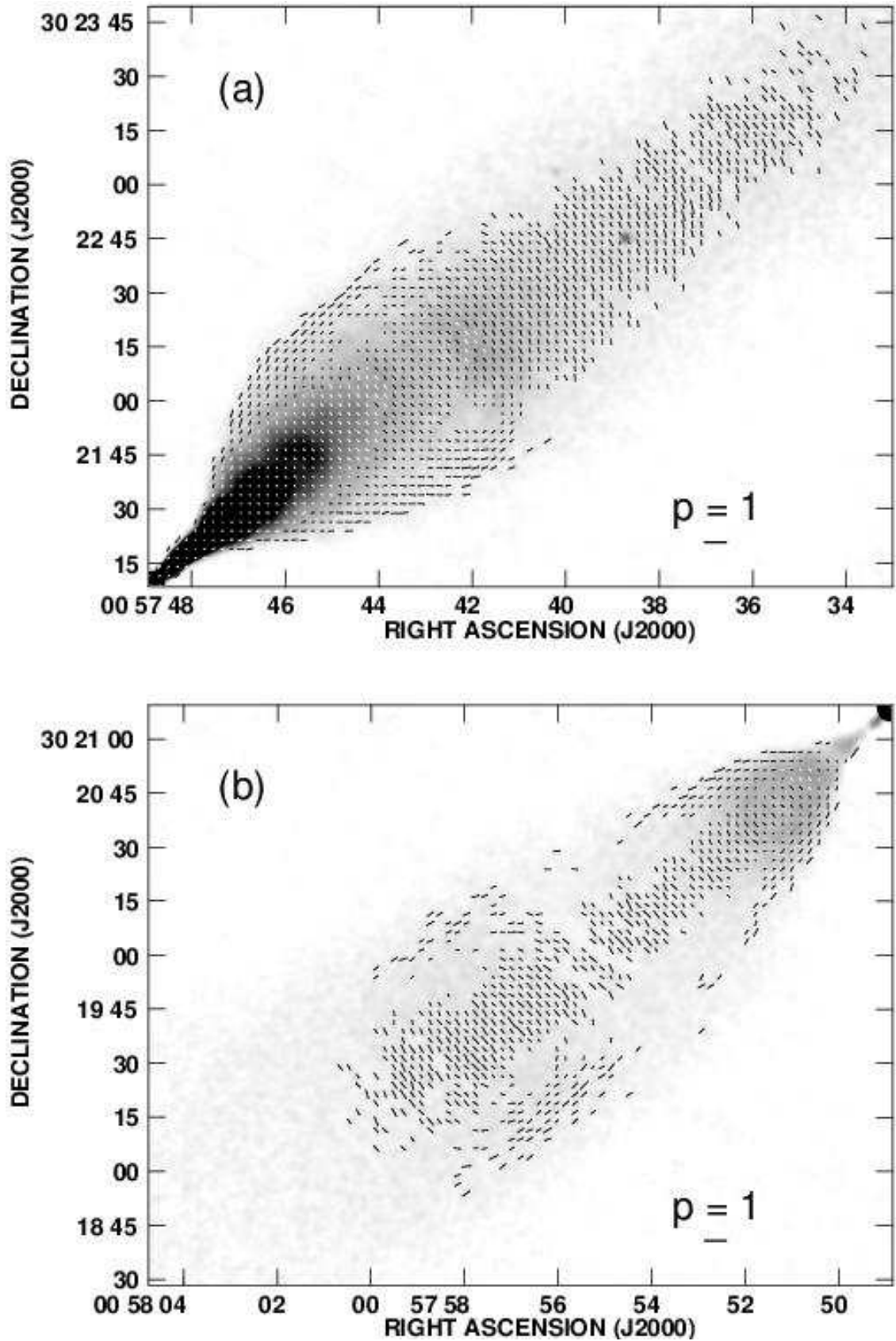}
\caption{Vectors whose lengths are proportional to the degree of polarization,
  $p$ at 4.985\,GHz and whose directions are those of the apparent magnetic
  field, derived by rotating the observed {\bf E}-vector position angles at that
  frequency by amounts derived from the 5-frequency RM fit of Section~\ref{RM}
  interpolated onto a finer grid. The vector scale is indicated by the labelled
  bars. The grey-scale is of total intensity at 4.985\,GHz and the resolution is
  2.35\,arcsec FWHM. (a) main jet; (b) counter-jet.
\label{fig:ivec2.35}}
\end{figure*}

Vectors whose magnitudes are proportional to $p$ at zero wavelength and whose directions
are those of the apparent magnetic field inferred from the rotation-measure fit
of Section~\ref{RM} are plotted in Fig.~\ref{fig:ivec5.5}. At higher resolution,
we derived the apparent field direction by interpolating the RM image onto a
finer grid and using it to correct the observed 5-GHz position angles
(Fig.~\ref{fig:ivec2.35}). The apparent field structure in the flaring region
(up to $\approx$70 arcsec from the nucleus) is
discussed extensively by \citet{CLBC}. Here, we concentrate on larger scales,
after the jets recollimate. 

\begin{figure}
\includegraphics[width=8.5cm]{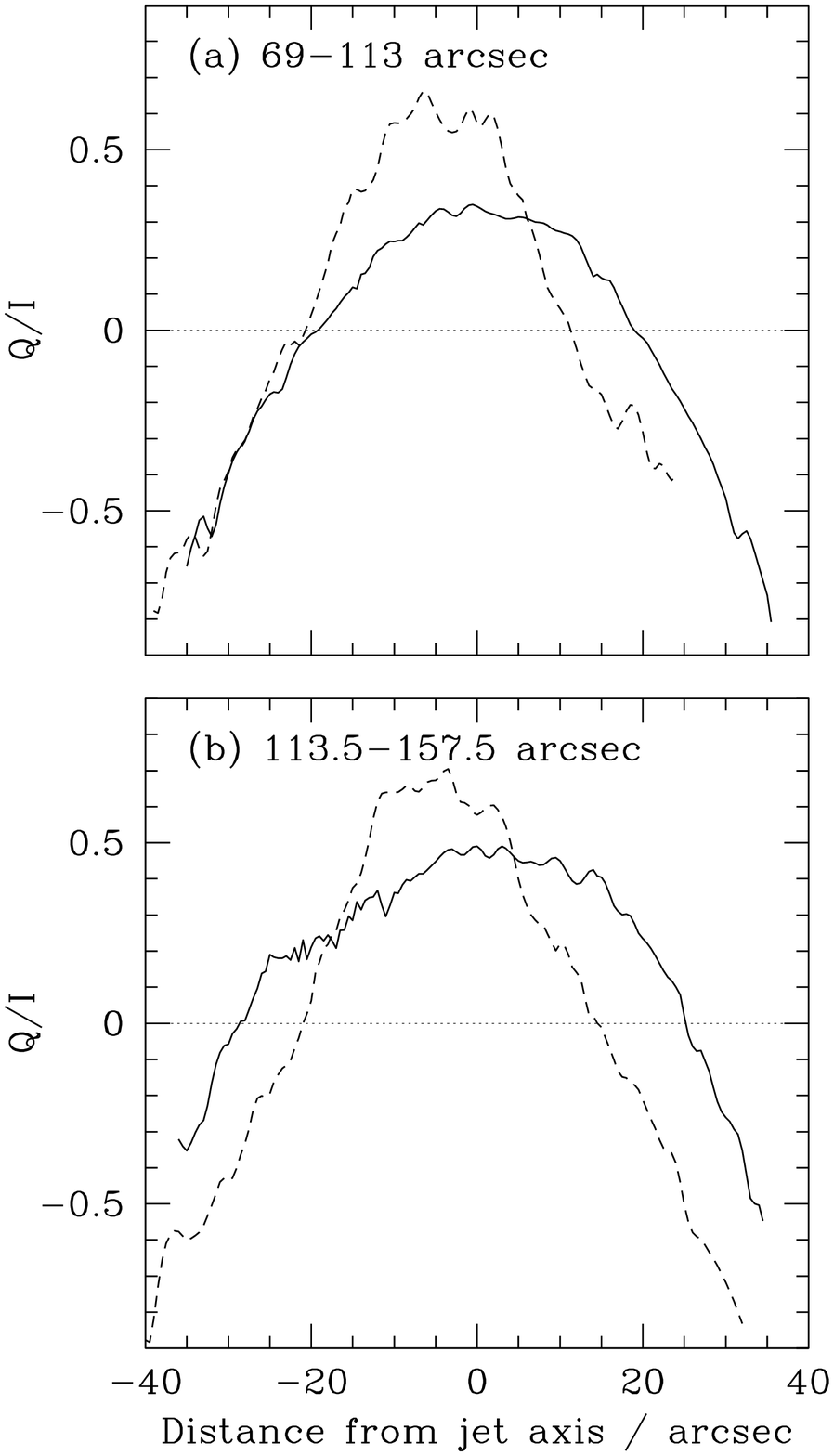}
\caption{Profiles of $Q/I$ for the main and counter-jets, integrated along the
  jet axis as described in the text. $|Q/I| = p$ for an apparent field direction
  either along or orthogonal to the jet, as seen here. $Q/I > 0$ corresponds to
  a transverse apparent field, $Q/I < 0$ to a longitudinal one. Full line: main
  jet; dashed line: counter-jet. (a) 69 -- 113 arcsec; (b) 113.5 -- 157.5 arcsec
  (the areas are the same as in Fig.~\ref{fig:transspec_long}).
  \label{fig:QIprof}}
\end{figure}

The signal-to-noise ratio for individual points, particularly at the jet edges,
is often $<$3 in linear polarization, causing them to be blanked in
Figs~\ref{fig:ivec5.5} and \ref{fig:ivec2.35}. In particular, it is impossible
to see whether the parallel-field edge continues to large distances in the main
jet. We therefore adopted the following procedure to derive the average degree of
polarization.
\begin{enumerate}
\item We first corrected the observed 5-GHz, 2.35-arcsec position angles for
  Faraday rotation using the linear model derived in Section~\ref{RM}, which is
  defined everywhere in the field, unlike the RM image.
\item We then changed the origin of position angle to be along the jet axis, so
  that apparent field along or orthogonal to the jet appears entirely in the $Q$
  Stokes parameter (also verifying that there is very little signal in $U$).
\item We then integrated $Q$ and $I$ along the jet axis from 69 -- 113 and from 113.5
  -- 157.5 arcsec from the nucleus (the same areas used for the profiles of
  spectral index in Fig.~\ref{fig:transspec_long}). Lack of short spacings at
  5\,GHz precludes extension of this analysis to larger distances. 
\item Finally, we divided the results to give transverse profiles of
  $Q/I$. Provided that the apparent field is either along or orthogonal to the
  jet axis, $p = |Q/I|$. We have chosen the sign convention so that $Q > 0$ for
  transverse apparent field and $< 0$ for longitudinal field.
\end{enumerate}
The resulting profiles are shown in Fig.~\ref{fig:QIprof}.
Figs~\ref{fig:ivec5.5} -- \ref{fig:QIprof} show that the apparent field
configuration found in the flaring region -- transverse on-axis and longitudinal
at the edges -- persists until at least 160\,arcsec in both jets. In particular,
the longitudinal-field edge of the main jet is easily detected in the profiles,
even though it is not clearly visible on the images. The main difference from
the corresponding profiles for the flaring region (fig.~9 of \citealt{CLBC}) is
that the on-axis polarization is higher in both jets at larger distances. This
is a continuation of the trend in the longitudinal profile shown by \citet[their
fig.~8]{CLBC}. As in the flaring region, the on-axis (perpendicular)
polarization is always higher in the counter-jet, reaching levels close to the
theoretical maximum of $p_0 = 0.69$ for the observed spectral index. In the main
jet, $p \approx 0.4$ on-axis. Both jets are very highly polarized at their
edges.

\citet{CLBC} modelled the three-dimensional structure of the field in the outer
parts of the flaring region as a mixture of toroidal and longitudinal components
of roughly equal magnitude on-axis but with the former dominant at the edge of
the jet. The differences between the two jets are attributed to relativistic
aberration, so the fact that they persist after the jets recollimate suggests
that there is little further deceleration in this region, despite the reversal
in sidedness at the edges of the jets (Section~\ref{Images}). A decrease in the
on-axis longitudinal field component, bringing the configuration closer to a
purely toroidal one, would result in a polarization distribution consistent with
that observed. Some longitudinal component must remain, however, otherwise $p$
would be close to $p_0$ on the axis for both jets. The very high degree of
polarization at the jet edge requires the radial field component to be very
small, as inferred for the flaring region by \cite{CLBC}.

\section{Summary}
\label{Summary}

We have imaged the jets in the nearby FR\,I radio galaxy NGC\,315 with the VLA
at five frequencies in the range 1.365 -- 5\,GHz and at resolutions ranging from
45 -- 1.5\,arcsec FWHM. Our total intensity observations reveal new details of
the structure, particularly around the sharp bend in the main jet.

The flaring regions of both jets, where they initially expand rapidly and then
recollimate, show a complex and previously unknown spectral structure. Within
15\,arcsec of the nucleus, the spectral index has a uniform value of $\alpha =$
0.61 in both jets. This region is associated with strong X-ray emission in the
main jet, high radio emissivity, complex filamentary structure, and fast flow
with $\beta \approx 0.9$ \citep{CLBC,WBH}.  Between 15 and 70\,arcsec, the
spectrum is steeper on-axis than at the edges of the jet. We have developed a
novel deprojection technique which allows us to isolate two spectral
components. The first (on-axis) forms a continuation of the jet base, its
spectral index flattening gradually from 0.61 to 0.55. The second (at the edge
of the jet) has $\alpha \approx$ 0.44 and is associated with a region where
strong shear is inferred \citep{CLBC}. We speculate that two different
acceleration mechanisms are involved, one associated with fast flow, dominant
close to the nucleus and capable of accelerating electrons to the very high
energies required to produce X-ray emission ($\gamma \sim 10^8$), the other
being driven by shear and generating the flatter spectral indices seen at the
edges of the jet.  Both mechanisms must efficiently generate the electrons with
$\gamma \sim 10^4$ which radiate at cm wavelengths. At distances
$\ga$70\,arcsec, the spectral index is consistent with $\alpha \approx 0.47$
everywhere.

We have imaged the variations of Faraday rotation over the jets. All of the
rotation is resolved and must originate mostly in foreground material. There is
no detectable depolarization. The largest contributions -- a constant term and a
linear gradient -- are probably Galactic in origin. We have also detected
residual fluctuations of $\approx$1 -- 2 rad\,m$^{-2}$ rms on scales $\sim$ 5 --
100\,arcsec.  The amplitude of fluctuations on scales $\ga$30\,arcsec is larger
by a factor $\approx$2 for the counter-jet, consistent with an origin in
magnetoionic material around the source, but not in the known X-ray-emitting
halo, whose core radius is too small. We model the Faraday-rotating medium as a
spherical halo with a core radius $\approx$225\,arcsec and derive an approximate
value for the product $(n_0/{\rm m}^{-3})^2 (B_0/{\rm nT})^2 (l/{\rm kpc})
\approx 700$, where $n_0$ is the central density, $B_0$ the central magnetic
field and $l$ is the magnetic-field correlation length. Our analysis is
therefore consistent with models of Faraday rotation proposed for rich clusters
(e.g.\ \citealt{CT}), but requires much lower densities and field strengths.  We
predict that a tenuous, group-scale halo should be detectable in sensitive X-ray
observations; measurement of its density will allow us to estimate the
magnetic-field strength.

We have derived the apparent magnetic field direction (corrected for Faraday
rotation) and degree of polarization at distances between 70 and 160\,arcsec
from the nucleus. The structure is qualitatively similar to that seen in the
flaring region, with transverse field on-axis and longitudinal field at the
edges of both jets, but the degree of polarization on-axis is larger.  The
difference in polarization structure between the main and counter-jets observed
in the flaring region by \cite{CLBC} persists at larger distances. This can be
explained fully as an effect of differential aberration on radiation from
intrinsically identical jets, as long as their velocities remain significantly
relativistic on the relevant scales.  The asymmetry in RM fluctuation amplitude is
consistent with the jet orientation required by this analysis and the presence
of a tenuous, magnetized group halo.

The large angular size of the flaring region in NGC\,315 and our use of deep
observations at several frequencies has allowed us to image spectral variations
at a level of detail not yet achieved in any other jet. Taken together with
X-ray imaging and modelling of the jet velocity field, this has given important
insights into the particle acceleration mechanisms. It will be interesting to see
whether our results apply to other FR\,I jets and to study the spectral
variations in NGC\,315 over a wider frequency range.

\section*{Acknowledgments}

JRC acknowledges a research studentship from the UK Particle Physics and
Astronomy Research Council (PPARC). The National Radio Astronomy Observatory is
a facility of the National Science Foundation operated under cooperative
agreement by Associated Universities, Inc. We thank Greg Taylor for the use of
his rotation-measure code, Karl-Heinz Mack for providing the 327-MHz WSRT image,
Frank Rieger for discussions on shear acceleration and the referee for helpful
comments. We also acknowledge the use of the {\sc healpix} package ({\tt
http://healpix.jpl.nasa.gov}) and the provision of the models of \citet{DK05} in
{\sc healpix} format.

\label{lastpage}
\end{document}